\documentclass[
journal=pasa,
manuscript=article, 
year=2023
]{cup-journal}

\pdfoutput=1

\usepackage{microtype,siunitx,booktabs}
\usepackage{longtable}
\usepackage{afterpage}
\usepackage{tabularx}
\usepackage{graphicx}
\usepackage{adjustbox}
\usepackage{supertabular}
\usepackage{amsmath}
\usepackage{caption}
\usepackage{subcaption}
\usepackage{amssymb}
\usepackage{tablefootnote}
\usepackage{capt-of}
\usepackage{mathtools}

\title{Azimuthal C/O Variations in a Planet-Forming Disk}

\author{Luke Keyte}
\affiliation{Department of Physics and Astronomy, University College London, Gower Street, WC1E 6BT London, United Kingdom}
\email[L. Keyte]{luke.keyte.18@ucl.ac.uk}

\author{Mihkel Kama}
\affiliation{Department of Physics and Astronomy, University College London, Gower Street, WC1E 6BT London, United Kingdom}
\alsoaffiliation{Tartu Observatory, Observatooriumi 1, T\~{o}ravere 61602, Tartu, Estonia}

\author{Alice S. Booth}
\affiliation{Leiden Observatory, Leiden University, 2300 RA Leiden, the Netherlands}

\author{Edwin A. Bergin}
\affiliation{Department of Astronomy, University of Michigan, 311 West Hall, 1085 S. University Ave, Ann Arbor, MI 48109, USA}

\author{L. Ilsedore Cleeves}
\affiliation{Department of Astronomy, University of Virginia, Charlottesville, VA 22904, USA}

\author{Ewine F. van Dishoeck}
\affiliation{Leiden Observatory, Leiden University, 2300 RA Leiden, the Netherlands}
\alsoaffiliation{Max-Planck Institut für Extraterrestrische Physik (MPE), Giessenbachstr. 1, 85748 Garching, Germany}

\author{Maria N. Drozdovskaya}
\affiliation{Center for Space and Habitability, Universität Bern, Gesellschaftsstrasse 6, CH-3012 Bern, Switzerland}

\author{Kenji Furuya}
\affiliation{National Astronomical Observatory of Japan, Osawa 2-21-1, Mitaka, Tokyo 181-8588, Japan}

\author{Jonathan Rawlings}
\affiliation{Department of Physics and Astronomy, University College London, Gower Street, WC1E 6BT London, United Kingdom}

\author{Oliver Shorttle}
\affiliation{Department of Earth Sciences \& Institute of Astronomy, University of Cambridge, United Kingdom}

\author{Catherine Walsh}
\affiliation{School of Physics and Astronomy, University of Leeds, Leeds, UK, LS2 9JT}

\accepted {03/03/2023}

\begin{document}

\newcommand{\msun}{$M_{\odot}$}
\newcommand{\mdisk}{$M_{\textrm{disk}}$}
\newcommand{\mdot}{$M^{todo}$}
\newcommand{\mstar}{$M_{\star}$}
\newcommand{\lsun}{$L_{\odot}$}
\newcommand{\lbol}{$L_{bol}$}
\newcommand{\lstar}{$L_{\star}$}
\newcommand{\rsol}{$R_{\odot}$}
\newcommand{\rstar}{$R_{\star}$}
\newcommand{\ngas}{$n_{\textrm{H2}}$}
\newcommand{\teff}{$T_{\textrm{eff}}$}
\newcommand{\rout}{$R_{\textrm{out}}$}
\newcommand{\mearth}{$M_{\oplus}$}
\newcommand{\mj}{$M_{J}$}
\newcommand{\sigmagas}{$\Sigma_\text{gas}$}
\newcommand{\sigmadust}{$\Sigma_\text{dust}$}
\newcommand{\deltadust}{$\delta_\text{dust}$}
\newcommand{\deltagas}{$\delta_\text{gas}$}
\newcommand{\gasdust}{$\Delta_\text{g/d}$}
\newcommand{\rsub}{$R_\text{sub}$}
\newcommand{\rgap}{$R_\text{gap}$}
\newcommand{\rcav}{$R_\text{cav}$}
\newcommand{\shgas}{[S/H]$_\text{gas}$}
\newcommand{\ohgas}{[O/H]$_\text{gas}$}
\newcommand{\chgas}{[C/H]$_\text{gas}$}
\newcommand{\kms}{km s$^{-1}$}

\newcommand{\luke}[1]{{\color{Maroon}{#1}\color{black}}}
\newcommand{\todo}[1]{{\color{blue}{[#1]}\color{black}}}
\newcommand{\mk}[1]{\color{magenta}{#1}\color{black}}


\begin{abstract}
The elemental carbon-to-oxygen ratio (C/O) in the atmosphere of a giant planet is a promising diagnostic of that planet's formation history in a protoplanetary disk. Alongside efforts in the exoplanet community to measure C/O in planetary atmospheres, observational and theoretical studies of disks are increasingly focused on understanding how the gas-phase C/O varies both with radial location and between disks. This is mostly tied to the icelines of major volatile carriers such as CO and H$_{2}$O. Using ALMA observations of CS and SO, we have unearthed evidence for an entirely novel type of C/O variation in the protoplanetary disk around HD~100546: an \emph{azimuthal} variation from a typical, oxygen-dominated ratio (C/O$\sim 0.5$) to a carbon-dominated ratio (C/O${\gtrsim}1.0$). We show that the spatial distribution and peculiar line kinematics of both CS and SO molecules can be well-explained by azimuthal variations in the C/O ratio. We propose a shadowing mechanism that could lead to such a chemical dichotomy. Our results imply that tracing the formation history of giant exoplanets using their atmospheric C/O ratios will need to take into account time-dependent azimuthal C/O variations in a planet's accretion zone.
\end{abstract}

Facilities such as the James Webb Space Telescope and ESA’s upcoming Ariel mission are opening a new era of exoplanetary studies, promising to characterise the atmospheric composition of over one thousand worlds by the early 2030s. Such observations provide exciting opportunities to constrain planetary formation and evolution models. To enable this science, we must understand the chemical budget of planetary nurseries.

Of particular significance are carbon and oxygen, which are among the most abundant volatile elements, and prominent in both protoplanetary disk and planetary chemistry. The elemental C/O ratio is a potential formation tracer, linking the volatile composition of an exoplanet atmosphere directly to the disk region in which it was accreted \citep{Madhusudhan_2012, Cridland_2016, Mordasini_2016}. The classical view of the C/O ratio in disk gas or solids is that of a multi-step function, with jumps at the snowlines of dominant molecular species \citep{oberg2011}. Considering dust growth and drift, vapour diffusion, refractory carbon destruction, chemical kinetics, and planetesimal and disk evolution adds substantial complexity, but the fundamental conclusion remains that the planetary C/O ratio is related to its formation history \citep{Madhusudhan_2014, Bergin_2016, booth_2017, Krijt2018, booth_ilee_2019, cridland_2019, vanthoff_2020, maps_7_bosman2021, turrini_2021, Van_Clepper_2022, hobbs_2022}.

Empirical findings based on spectroscopic observations support the hypothesis that the gas-phase elemental C and O abundances can vary both with radial location in a disk and between different disks \citep{favre_2013, qi_2013,  vandermarel_2016, Du_2017, cleeves2018, zhang_2019, Zhangetal2020}. While determining the precise C/O ratio in planet-forming gas is generally not simple, distinguishing between oxygen- or carbon-dominated chemistry (C/O\,$<1$ or $\gtrsim 1$, respectively) is more feasible. The two chemical regimes are characterised by entirely different compositions. The C$_2$H molecule is a powerful C/O tracer, strong C$_2$H emission being a beacon of C/O\,$\gtrsim1$ in the disk gas due to the orders-of-magnitude increase in the abundance of hydrocarbons when excess carbon is available \citep{Bergin_2016, miotello_2019, bergner_2019, maps_7_bosman2021}. Another probe is the CS/SO molecular ratio, which varies by up to four orders of magnitude for small variations in C/O \citep{Semenov_2018, Booth_2021}, and has led to claims of C/O\,$>1$ in at least six disks \citep{dutrey_2011, maps_7_bosman2021, maps_12_legal2021}. Such results are usually disk-averaged (i.e. unresolved) or indicative of \emph{radial} variations in the gas composition. Until now, combined observations of CS and SO have not been made in any disk, so it has not been possible to fully utilize the diagnostic power of the CS/SO ratio in cases of asymmetric emission.

In this paper, using observations of both CS and SO, we present a remarkable \emph{azimuthal }C/O variation in the planet-forming disk around HD\,100546, the first time such a variation has been observed.


\section{RESULTS}
\label{sec:results}

\subsection{Data \& observational findings}
\label{subsec:results_intensitymaps}

\begin{figure}[!t]
\centering
\includegraphics[clip=,width=1\linewidth]{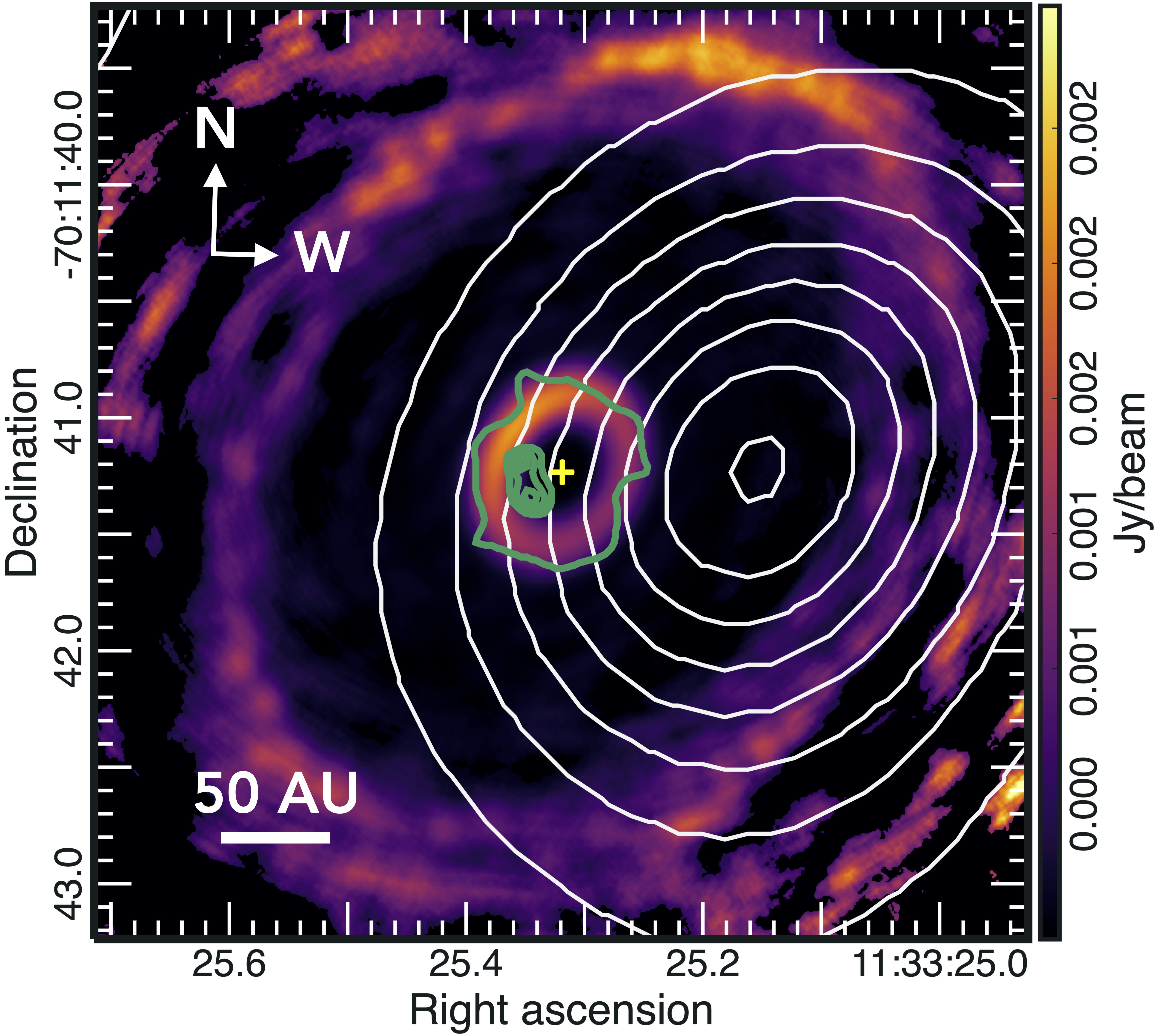}
\caption{\textbf{Azimuthal disparity of SO and CS emission} ALMA 870$\mu$m continuum emission map \citep{fedele_2021}, overlaid with SO 7$_7$-6$_6$+7$_8$-6$_7$ emission contours (green) and CS 7-6 emission contours (white). The continuum emission has been scaled by $r^3$ to highlight emission in the outer disk. Contours are logarithmically spaced between $1\sigma$ and peak flux. The position of the star is denoted by the yellow cross. Beam sizes are $\sim$0.18"/$20$ au (SO) and $\sim$4.78"/$525$ au (CS).}
\label{fig_contours}
\end{figure}

\begin{figure*}[hbt!]
\centering
\includegraphics[width=\textwidth]{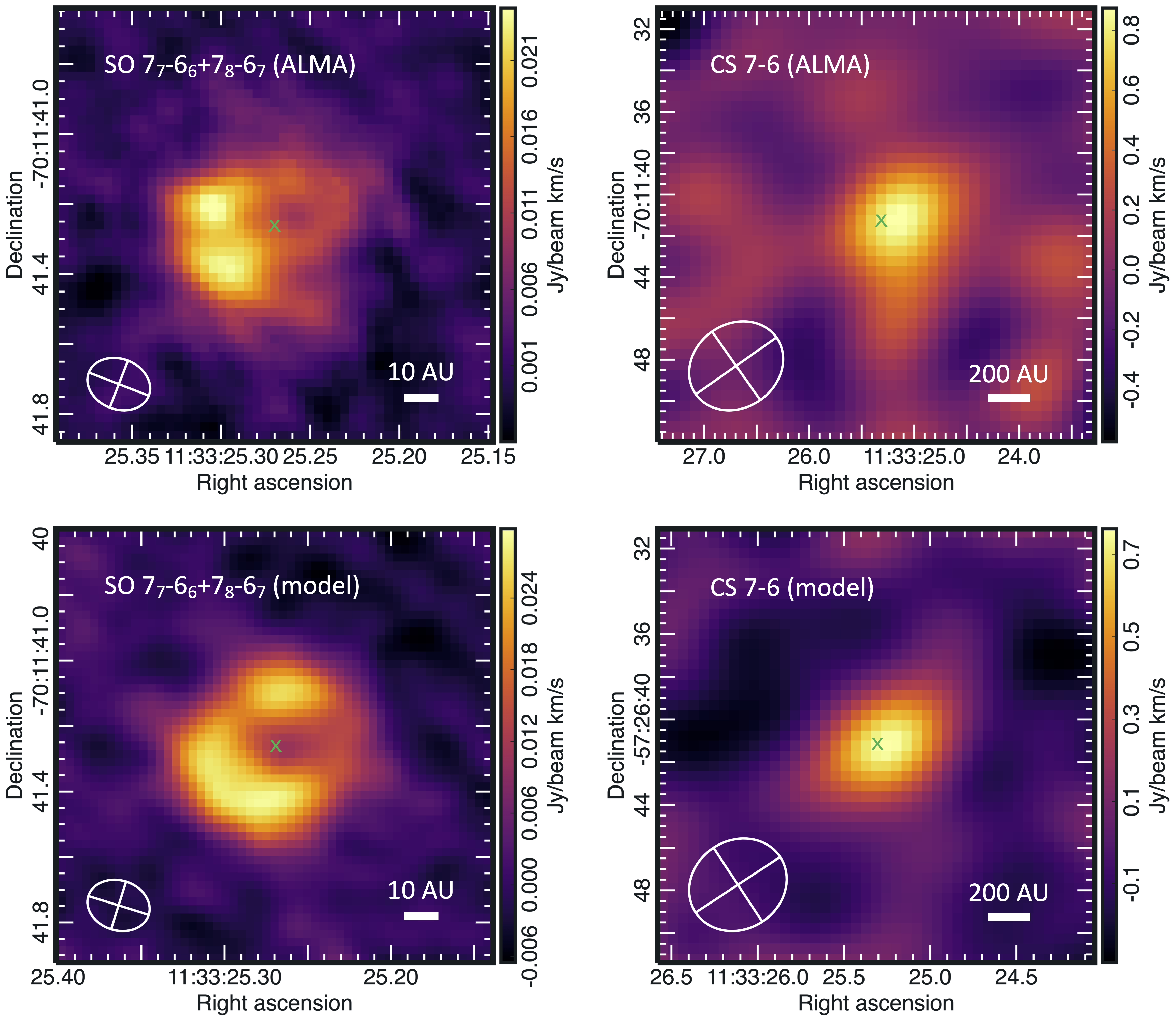}
\caption{\textbf{Detected and modelled SO and CS emission in HD~100546}. \emph{Top left: }Integrated intensity map of the stacked SO $7_7-6_6$ and $7_8-6_7$ transitions observed with the ALMA (\citep{booth_2022}). \emph{Top right: }Integrated intensity map of the CS 7-6 transition observed with ACA, with a $1\sigma$ clip. \emph{Bottom left: }Modelled stacked SO $7_7-6_6$ and $7_8-6_7$ integrated intensity map. \emph{Bottom right: }Modelled CS 7-6 integrated intensity map. The position of the star is indicated with the green `x'.}
\label{fig_moment0}
\end{figure*}

We present a new detection of CS in HD\,100546, alongside ALMA SO observations first presented in \citep{booth_2022} (Figures \ref{fig_contours} and \ref{fig_moment0}). HD\,100546 is a well-studied $2.49 \pm 0.02$ \msun \ Herbig Be star, with an estimated age of $\sim5$ Myr \citep{Arun_2019}, and distance of $110 \pm 1$~pc. The star hosts a bright disk with a central dust cavity out to $13\,$au and another dust gap between $r \sim 40$ - $150\,$ au, bounded on both sides by dust rings \citep{Walsh_2014, fedele_2021}. Multiple observations provide evidence for at least two planetary candidates within the disk at $r \sim 13$ and $55$ au \citep{Walsh_2014, quanz_2015, Currie_2015, pinilla_2015}.

CS is detected at $9\sigma$ confidence ($0.90\pm0.10$\,Jy\,beam$^{-1}$\,km s$^{-1}$ at emission peak, as measured from the integrated intensity map), using the Atacama Compact Array (ACA) in Cycle\,4 (Figure \ref{fig_moment0}, top right). The emission is essentially unresolved, since the beam size (4.78") is almost equal to the radial extent of the gas disk, as traced by CO \citep{Walsh2017}. Fitting a Gaussian, we find the peak of the CS emission to be significantly offset from the host star by $\sim 1$". We confirm that the offset is related to a physical characteristic of the source, rather than a pointing error (\ref{sec:appendix_continuum_c18o}). We determine the radial separation between the peak of the emission and the host star by exploiting the known inclination and position angle of the disk to deproject the image, finding the emission peak to be radially offset $\sim 100$ au from the source. \looseness=-1

We complement the new CS detection with high resolution Cycle\,7 observations of SO, first presented in \citep{booth_2022} (Figure \ref{fig_moment0}, top left). The emission primarily emanates from the inner dust cavity and the inside edge of the dust ring ($r\sim 13$ au), displaying a clear azimuthal brightness asymmetry, where emission from the eastern side of the disk is a factor of $\sim 2$ brighter than from the western side.

\begin{figure*}[hbt!]
\centering
\includegraphics[width=\textwidth]{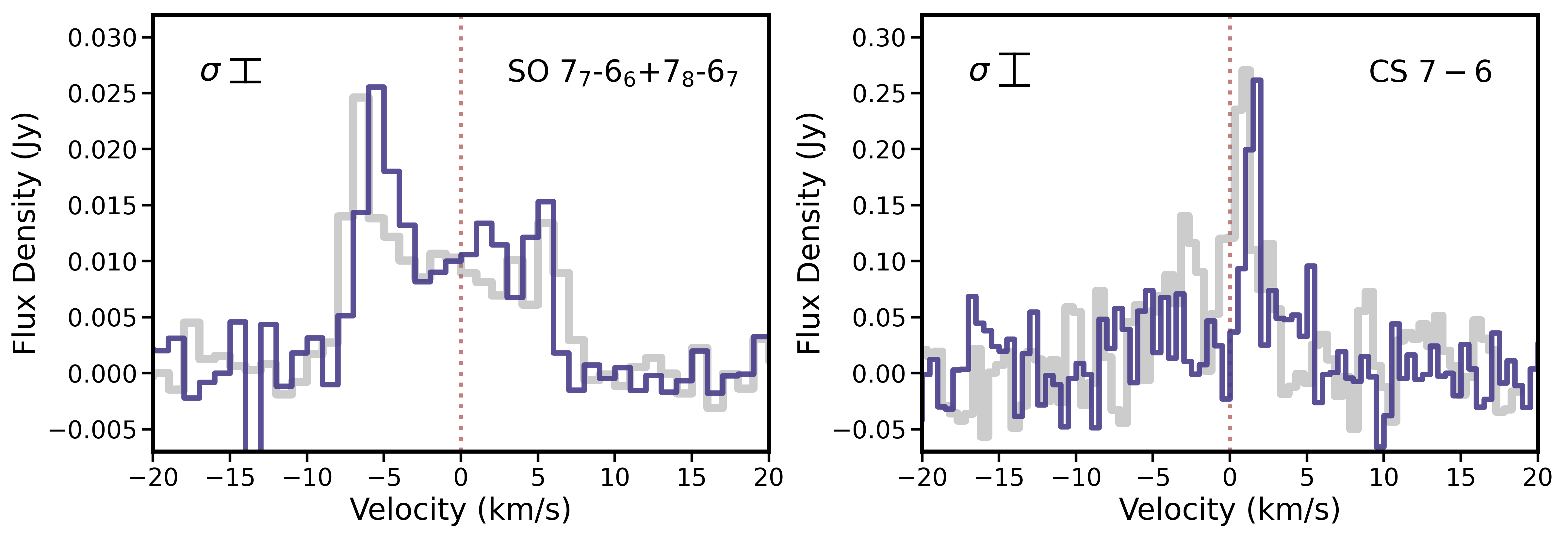}
\caption{\textbf{Spectral line profiles}. \emph{Left:} Observed stacked SO $7_7-6_6$ and $7_8-6_7$ spectrum extracted using a 0.6" elliptical mask (grey) and model (blue). \emph{Right:} Observed CS 7-6 spectrum extracted using a 5" elliptical mask smoothed by the beam (grey) and model (blue). Line profile velocities have been corrected for the source velocity ($V_\text{LSRK} = 5.7 \text{ km s}^{-1}$).}
\label{fig_spectra_cs_so}
\end{figure*}

The distinct morphologies of the SO and CS emission are mirrored in their respective spectral lines (Figure \ref{fig_spectra_cs_so}), which have unusual and disparate velocity profiles. The SO line is broad ($\text{FWZI}\sim 15$ km s$^{-1}$) and asymmetric, with a prominent blueshifted Keplerian peak at $\sim -7.5$ km s$^{-1}$. In contrast, the CS emission is narrower (FWZI $\sim 10$ km s$^{-1}$) and sharply peaked in the red ($\sim +1.5$ km s$^{-1}$). \looseness=-1

In summary, the CS and SO emission display clear azimuthal asymmetries in their spatial morphologies, and peculiar spectral line profiles. It is striking that emission from each of these species appears to emanate from distinct and opposite azimuthal regions of the disk. We argue that these features can be fully explained by chemistry resulting from azimuthal C/O variations in the disk.

\subsection{Modelling}
\label{subsec:results_modelling}

To investigate the origin of the spatial and spectral asymmetries in the CS and SO emission, we ran source-specific models using the 2D physical-chemical code DALI \citep{Bruderer2012, Bruderer2013}. The disk chemical composition is obtained from a chemical network simulation, in which the gas-grain chemistry is solved time-dependently. Our model uses a geometry outlined in Figure \ref{fig_disk_schematic}, in which the disk is composed of two chemically distinct regions. The majority of the disk has a composition consistent with previous studies of HD~100546, in which C/O=0.5 \citep{Kama2016b}. We vary the composition in a small angular region of the disk, such that C/O is elevated within an azimuthally localized `wedge' (C/O>1), dictated by variations in the gas-phase H$_2$O, CO, and atomic O abundances (see \ref{sec:appendix_timescales}).

We explored a wide parameter space, taking into account a range of wedge sizes and azimuthal locations, carbon and oxygen abundances, and chemical timescales. The model presented here incorporates a high-C/O wedge extending azimuthally $60^\text{o}$, centered ten degrees north of west. Modelled integrated intensity maps for both the CS 7-6 and stacked SO 7$_7$-6$_6$ + 7$_8$-6$_7$ emission are presented in Figure \ref{fig_moment0} (lower panel). Our model reproduces the CS emission morphology well. The emission peak is significantly offset towards the west ($\sim 75$ au deprojected), with a peak flux that matches the observation within a factor of $\sim 1.1$. The brightness asymmetry in the SO emission is also reproduced by our model, peaking towards the southwest at a radial separation of $\sim 20$ au. The peak flux matches the observations within a factor of $\sim 1.2$. A more refined model may be needed to fully reproduce the SO brightness distribution, although we note that the precise distribution can vary depending on the parameters used for the data reduction \citep{booth_2022}. The east/west asymmetry is always maintained, and is the key feature reproduced by our model. \looseness=-1

The modelled spectral lines are shown in Figure \ref{fig_spectra_cs_so}. The modelled CS line profile includes a prominent narrow single peak, red-shifted $\sim 2$ km s$^{-1}$ from the source velocity, matching the observation to within $\sim 0.5$ km s$^{-1}$. While it is possible to match the peak location more precisely by changing the orientation of the high-C/O wedge, this results in a slightly lower peak flux (\ref{sec:appendix_wedge_variations}). The modelled SO line profile reproduces both the double-peaked structure and peak flux value. The linewidth is a close match to the observation, where the flux density in the blue shifted component is $\sim 1.5 \times$ greater than that of the red-shifted component. 

Modelled CS and SO abundance maps are presented in \ref{sec:appendix_cbfs}. We note that our model predicts that the bulk of the SO emission emanates from the dust ring just outside the cavity, whereas \citep{booth_2022} used the line kinematics to infer that it originates primarily from within the cavity itself. One possible explanation is that our model lacks a smooth transition between the cavity and dust ring, instead having a sharp boundary. Additionally, the inclination and stellar mass used in our model differs from the values reported in \citep{booth_2022}. Literature values vary between $i=32-44^\circ$ \citep{Walsh_2014, pineda_2019} and $M_*=2.2-2.5 \; M_\odot$ \citep{pineda_2019, Arun_2019, wichittanakom_2020} which can result in variations of the predicted inner edge of the SO emission ($\sim9-18$ au).

\section{DISCUSSION}
\label{sec:discussion}

We have shown that the emission from CS and SO in the HD~100546 protoplanetary disk can be well-reproduced by a chemical model that incorporates an azimuthal C/O variation. This adds a new dimension of complexity to the relationship between C/O in disks and their planetary progeny. We now aim to understand the origin of this novel type of chemical dichotomy. \looseness=-1

The depletion of volatile elemental carbon and oxygen is ubiquitous in protoplanetary disks around both T Tauri and Herbig Ae/Be stars. For instance, disks around AS~209, MWC~480, and HD~163296 all exhibit sub-stellar C/H and O/H ratios \citep{maps_7_bosman2021}. Oxygen is typically more depleted than carbon due to its removal from the disk atmosphere through the freezing out of water onto large dust grains, resulting in elevated C/O ratios ($\sim 2$). However, our modelling of HD 100546 finds no evidence of significant oxygen depletion relative to carbon in this disk, with a best-fit \emph{disk-averaged} C/O ratio $<1$. The majority of the disk is warm enough to preclude freeze-out of CO and CO$_2$ (\ref{sec:appendix_tempdens}), and water loss through freeze-out onto large grains is tempered by the presence of a dust cavity where large grains are heavily depleted (small grains are not as significant for \emph{permanent} freeze-out, as they cycle vertically within the disk due to turbulence, releasing their ices upon each return to the disk atmosphere). Photodesorption also limits the extent of the water snow surface (Figure \ref{fig_tdust_snowsurface}) Analysis of water emission from HD 100546 reveals a high H$_2$O abundance in the photodesorption region is necessary to match observations, with line kinematics indicating that the emission extends out to r$\sim300$ au \citep{pirovano_2022, vandishoeck_2021}. Therefore, we expect the majority of the HD~100546 disk to have a gas-phase C/O ratio closer to 0.5, and attribute the observed asymmetries to a region of elevated C/O localized in azimuth. \looseness=-1

\begin{figure*}[t]
\centering
\includegraphics[clip=,width=1\linewidth]{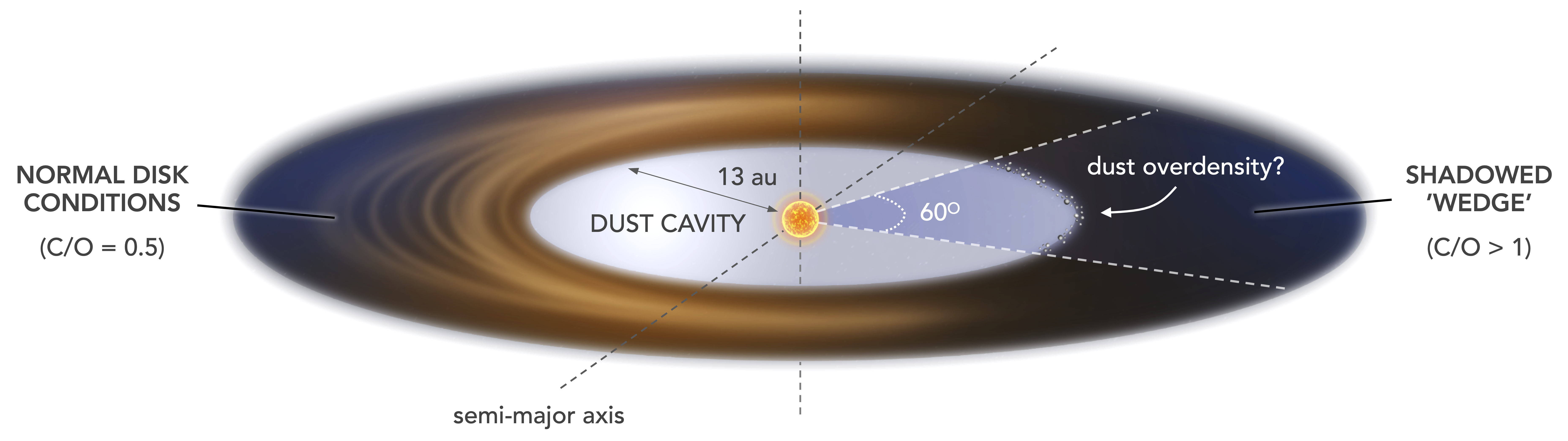}
\caption{\textbf{Geometry of the HD~100546 disk model. } The disk is composed of two chemically distinct regions; the majority of the disk has C/O=0.5, apart from a 60$^\circ$ arc where C/O>1 We propose that an overdensity of dust, associated with a newly forming planet within the cavity, casts a shadow over an azimuthally localized region of the disk. This results in lower temperatures, which causes additional H$_2$O freeze-out, leading to an elevated gas-phase C/O ratio. Note that this schematic is not to scale; the gas disk extends out to $\sim 500$ au.}
\label{fig_disk_schematic}
\end{figure*}

The main feature we have identified is an azimuthally confined zone of elevated CS abundance, coincident with a region of depleted SO, which we ascribe to a local enhancement in the C/O ratio ($>1$). How could this come about? Asymmetries in the structure of both dust and gas are common in protoplanetary disks, particularly transition disks like HD~100546 which have large central cavities \citep{Francis_vdM_2020}. Dust asymmetries are often attributed to the trapping of millimeter-sized grains in vortices formed by the Rossby Wave Instability \citep{Lovelace_1999}, induced by a planetary companion. Vortices can also form through various hydrodynamical instabilities, or at the edges of low-viscosity 'dead-zones'. In recent years, high-resolution ALMA observations have enabled comparisons between dust asymmetries and molecular gas at small scales. While several studies have drawn tentative links \citep{maps_3_law2021, maps5_zhang2021, maps6_guzman2021, maps8_alarcon2021, maps9_ilee2021, vandermarel_2021}, there is often no clear connection to be made. Furthermore, asymmetries observed in a particular species are often not observed in other species or transitions of the same species within the same disk. The physical mechanisms responsible for gas asymmetries cannot therefore be easily attributed to vortices.

In HD~100546, gas-phase asymmetries have previously been observed in a range of molecular species, including various CO transitions \citep[e.g.][]{panic2010, Kama2016b, miley_2019}, OH \citep{fedele_2015}, and SO \citep{booth_2017}. These have often been attributed to temperature variations which are thought to result from obscuration by a warped inner dust disk \citep{panic2010, Walsh2017}, such that one side of the disk is 10-20 K cooler than the other. In such a scenario, an azimuthal variation in the temperature structure could have significant impact on the disk chemistry, resulting in azimuthal variations in molecular abundances \citep{Young_2021}. However, near-IR observations at small spatial scales ($\sim$1 au) find no evidence supporting an inclined inner dust disk ($r<1$ au) \citep{Garufi_2016, Follette_2017, Lazareff_2017, bohn_2021}, which may indicate that the structure of the gas and dust is significantly different within the inner few au. Currently, no physical mechanism is known to cause such decoupling between the gas and small dust grains, suggesting that another process might be at play. \looseness=-1

An alternative scenario is that asymmetric emission is connected to on-going planet formation within HD\,100546's inner cavity. 
\citep{booth_2022} propose that the observed SO asymmetry may be tracing shocked gas in the vicinity of a circumplanetary disk. Indeed, the peak of the observed SO emission is cospatial with the location of a protoplanet candidate inferred from scattered light images \citep{Currie_2015} and excess CO emission \citep{brittain_2019}. Comparing Cycle 7 and Cycle 0 spectra of SO emission provides further evidence of a newly-forming planet, as the shift in emission peak is consistent with a hot-spot of molecular gas in  orbital rotation within the inner cavity \citep{booth_2017, booth_2022}. Our findings do not rule out this possibility. While our model can be modified to account for an additional component of SO emission related to a CPD, a CPD alone is not able to account for the asymmetries observed in both the SO and CS. Both the kinematics and the spatially resolved SO emission are consistent with a Keplerian protoplanetary disk. The spatial distribution of the emission indicates that any contribution from a CPD to the total SO flux would be relatively small.

This leads us to propose an alternative explanation, consistent with an azimuthal C/O variation. We suggest that an overdensity of dust associated with the inner protoplanet casts a shadow on an azimuthally localised region of the outer disk. This causes dust temperatures in the disk atmosphere to decrease, which leads to additional H$_2$O freeze-out on grain surfaces. In turn, this locks a significant fraction of gas-phase oxygen into ices, causing the local gas-phase C/O ratio to become super-solar. As the disk chemistry rebalances, SO is rapidly destroyed while CS is rapidly formed, on timescales shorter that the shadow transit time. \looseness=-1

Falsifying this hypothesis using past observations is not straightforward. Our understanding of dust substructures within HD~100546 is shaped by both infrared and (sub)millimeter observations, which contain many features. These include spiral arms, dark and bright azimuthal wedges, emission hotspots  \citep{Garufi_2016, sissa_2018}. and an inner ring sculpted by a maze of ridges and trenches \citep{perez_2020, fedele_2021}. We note that the location of the protoplanet within HD 100546’s inner cavity is not coincident with our high-C/O wedge towards the west, but closer to the region of bright SO emission. Nevertheless, material associated with a forming planet can be highly azimuthally and vertically extended \citep{zhu_2014}. Indeed, several features suggestive of shadowing have been identified on the western side of the disk. The most prominent is a dark region towards the northwest, already linked to a possible large-scale shadow \citep{Garufi_2016}. This region covers a similar azimuthal width to the high-C/O wedge used in our model, and overlaps with it significantly (the local minimum in the azimuthal brightness profile is orientated slightly further northwest by $\sim 10^\circ$). At least one other similar dark wedge was identified on the opposite side of the disk \citep{norfolk_2022}. Other features that could be related to a dust overdensity include a horseshoe-like structure identified in 7mm observations at the inner southwest edge\citep{wright_2015}, and a bar-like structure seen in H$\alpha$ polarized light, which may be a ``streamer'' of dust dragged in by gas flowing from the outer to inner disk \citep{mendigutia_2017}. However, it is unclear whether such features could lead to large-scale shadowing. So, while HD~100546 clearly displays a number of morphological features that could be related to shadowing, linking any such feature to the region of high-C/O identified here remains open to interpretation.

To determine if a shadow could cause the required chemical changes to produce C/O > 1, we examined three timescales: cooling, freeze-out, and chemical conversion (\ref{sec:appendix_timescales}). As the shadow falls on a part of the disk, the dust must cool enough for H$_2$O to freeze out. Kinetics will then funnel O from other reservoirs (atomic O, CO) into the water-ice sink, eventually elevating the CS abundance and depleting SO within the shadowed region. The combined time needed for these processes to operate must be shorter than the time spent in shadow, which converges to $\sim5$ years in the outer disk (based on model parameters). 
Considering a range physical conditions from our model in the CS-emitting region (\ref{sec:appendix_cbfs}), we find that the C/O ratio can exceed unity in $\leq 5$ years inside of $r \lesssim 200$ au. These results are illustrated in Figure \ref{fig_timescales}. The shadowing hypothesis can be tested in the near future with high-resolution observations of the CS emission, to establish whether the feature moves. Moving shadows of this nature have already been observed in other disks such as TW Hya \citep{debes_2017}.

\begin{figure}[!t]
\centering
\includegraphics[clip=,width=1\linewidth]{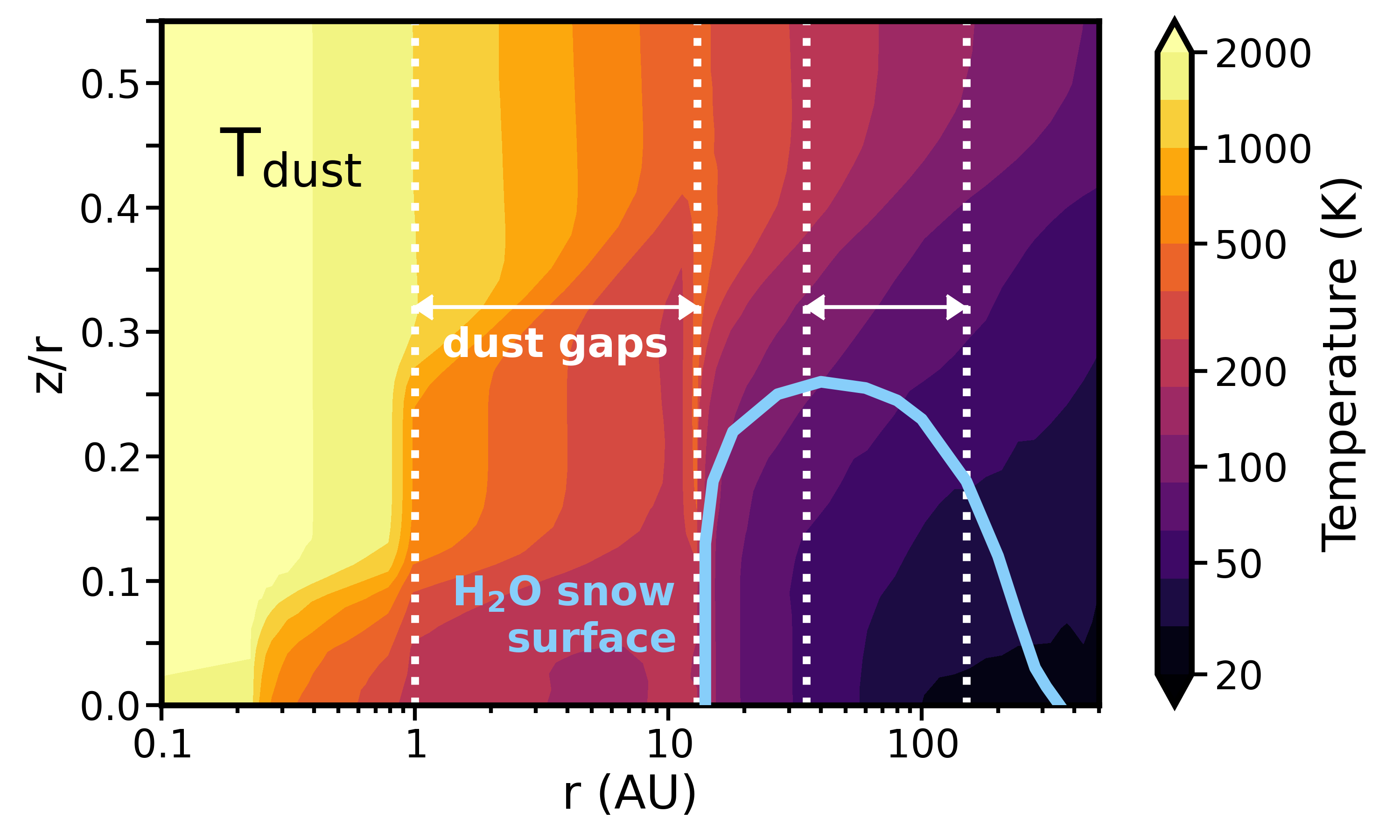}
\caption{\textbf{HD~100546 model dust temperature map}. Most of the disk is warm enough to preclude freeze-out of CO and CO$_2$. The H$_2$O snow surface (blue line) largely coincides with a dust gap in the outer disk (white dotted lines). Millimeter-sized grains are highly depleted in this region, removing grain surface area available for permanent H$_2$O freeze-out. The outer edge of the snow surface is curtailed by photodesorption.}
\label{fig_tdust_snowsurface}
\end{figure}

Regardless of the precise mechanisms which lead to azimuthal C/O variations in the HD 100546 disk, it is clear that such a chemical dichotomy will have a profound impact on the final composition of planets forming within it. Growing planets are supplied by gas and dust from their surroundings. Planets which move in and out of two chemically distinct regions during the course of their evolution can be expected to have chemically complex envelopes, formed of material accreted from both regions. The degree to which envelope composition mirrors that of either region in the disk will be governed by complex chemical and physical processes. If shadowing is indeed responsible for the azimuthal C/O variation, we may expect its effect on planetary composition to be even more profound in warm disks such as HD 100546, where there is a higher fraction of gas-phase water available for freeze-out. \looseness=-1

The results presented here therefore add a new consideration to the way in which we interpret observations and model gas-phase asymmetries in protoplanetary disks. The classical view of a radially varying C/O ratio must be readdressed if we are to draw meaningful links between the composition of exoplanet atmospheres, and the disks in which they form. Determining the C/O ratio at small spatial scales must be a major goal of future observations, if we are to build models that can meaningfully predict planetary formation pathways.


\small{

\section{METHODS}
\label{sec:methods}

\subsection{Data reduction}
\label{subsec:data_reduction}

HD~100546 was observed with the Atacama Compact Array (ACA) in Band 7 during Cycle 4, in two separate execution blocks on November 14th and November 24th 2016 (program 2016.1.01339.S, PI: M. Kama). The observations cover 8 molecular rotational transitions of 7 sulfur-bearing species/isotopologues, outlined in Table \ref{table:aca_observations}. Eight scans were performed for a total on-source time of 50.64 minutes, with baselines ranging from 8-45 m. System temperatures varied from 103-170 K and the average precipitable water vapour was 1.0 mm. J1058+0133 was used as both the bandpass and flux calibrator, while J1147-6753 was used as the phase calibrator.

The data reduction was completed using the ALMA Pipeline in the Common Astronomy Software Package (CASA) version 5.6.1-8. Self-calibration was performed but found to have marginal impact due to low S/N of the data. Continuum and line imaging were performed with the tCLEAN algorithm using natural weighting in order to maximize the S/N ratio of the data. The resulting synthesized beam size was $\sim$ 4.78" x 4.06", with slight variations depending on the spectral window. We used a cell size of 0.5" to ensure that the beam is well sampled. Continuum subtraction was performed with the CASA task \emph{uvcontsub}, using a single-order polynomial fit to the line free channels. The spectral resolution, bandwidth, synthesized beam size for each of the transitions are listed in Table \ref{table:aca_observations}. 

CS 7-6 at 342.883 GHz is successfully detected, while all other transitions are undetected. The CS 7-6 integrated intensity map (Figure \ref{fig_moment0}, top right) was generated from a 20"x20" region centred on the source, where the integrated intensity is determined between -14.5 to 25.5 \kms, corresponding to channels expected to contain significant emission ($\sim \pm 20$ km s$^{-1}$ from the source velocity $V_\text{LSRK} = 5.7$ km s$^{-1}$). The detection is made at a 9$\sigma$ confidence level, with a peak flux of 0.90 Jy beam$^{-1}$ km s$^{-1}$, and rms of 0.10 Jy beam$^{-1}$ km s$^{-1}$ as measured from the emission-free regions of the integrated intensity map. Channel maps are presented in \ref{sec:appendix_channel_maps}, from which we measure a peak flux density of 0.21 Jy beam$^{-1}$ and rms of 0.024 Jy beam$^{-1}$.
\looseness=-1

We extracted the spectrum from the CLEAN cube using an elliptical aperture with a 5.0" radius centred on the source (approximately the same size as the disk as traced by $^{12}$CO emission \citep{Walsh_2014}), where the edges of the mask were smoothed by the beam. We also extracted a spectrum using a Keplerian mask, which excludes noisy pixels that are not directly associated with emission from a disk in Keplerian rotation \citep{teague2020}. The mask identifies which pixels in the image cubes have Doppler shifted line velocities that match the Keplerian velocity, based on the velocity profile of a disk rotating around a star of mass \mstar$=2.4$ \msun. Pixels with velocities that do not match the Keplerian velocity are masked. The mask is convolved with a beam of equal size to the observation, in order to provide a buffer between the mask edge and emission edge. The total disk-integrated flux was extracted using the CASA task \emph{specflux}, and determined to be 0.62 Jy km s$^{-1}$ for the Keplerian-masked cube, where the mask was cut at $\pm 4$ km s$^{-1}$ either side of the source velocity in order to remove noisy channels. The disk-integrated flux extracted from the elliptically masked cube was 1.02 Jy km s$^{-1}$. Due to the peculiar nature of the CS line profile, our analysis utilises only the elliptically-masked spectrum, in order to not mistakenly remove any real emission.

\begin{figure}[ht!]
\centering
\includegraphics[clip=,width=0.99\linewidth]{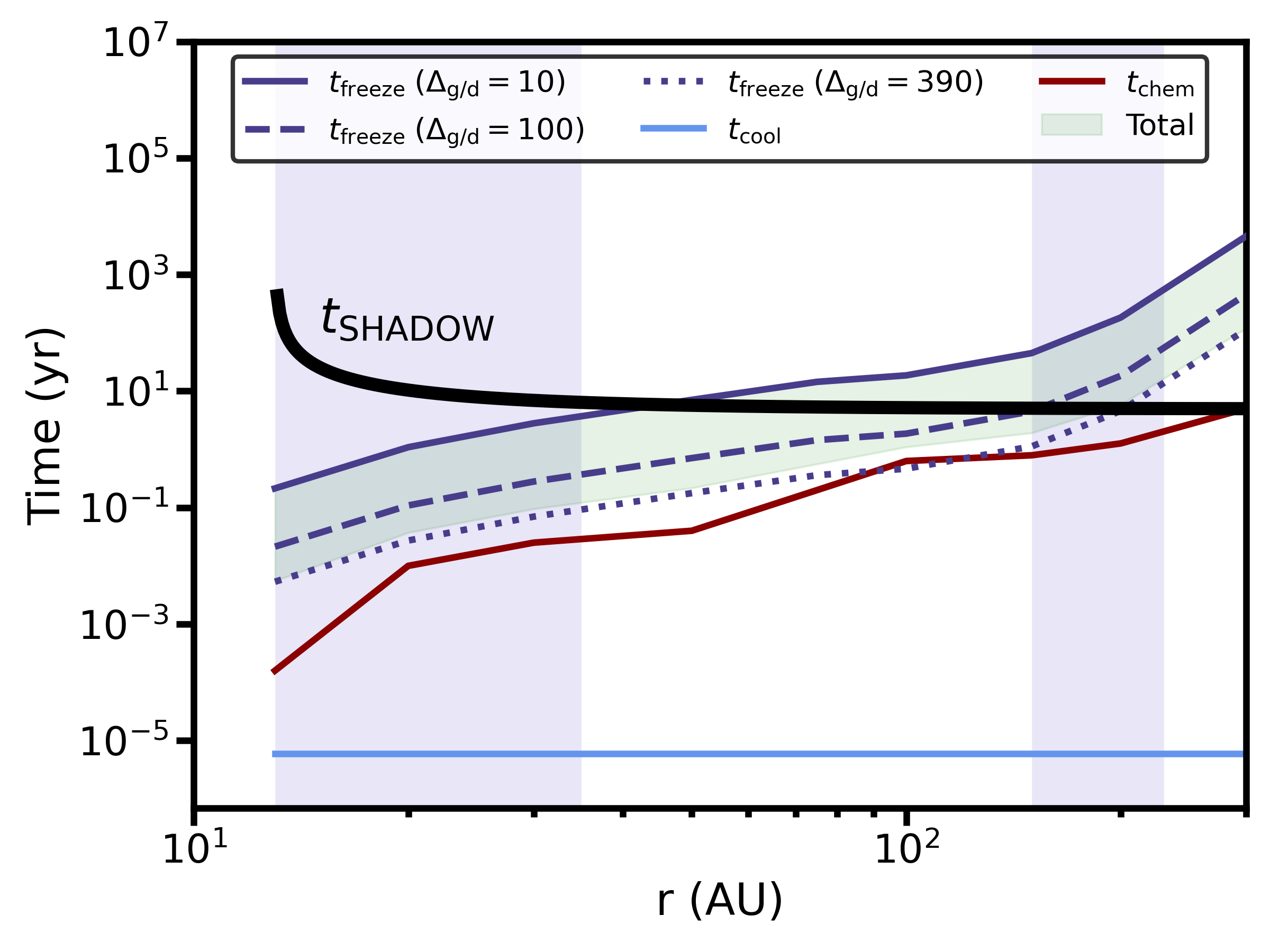}
\caption{\textbf{Cooling, freeze-out, and chemical timescales}. The maximum time available for the combination of cooling, freeze-out, and shadowing to occur is dictated by the period a specific region of the disk remains in shadow, which is related to the orbital period of the shadowing material (black line). The cooling timescale is denoted in light blue, freeze-out timescale in  dark blue (for a range of $\Delta_\text{g/d}$), and chemical timescale in red. The green shaded area represents the sum total of these timescales within the range of considered $\Delta_\text{g/d}$, which follows the freeze-out timescale since the cooling and chemical timescales are negligable. Purple shaded regions denote location of millimeter dust rings.}
\label{fig_timescales}
\end{figure}

We employed a number of techniques in an attempt to extract weak spectral signatures from the remaining spectral windows centred on other molecular species (Table \ref{table:aca_observations}). We began by extracting spectra from the CLEAN image cubes using an elliptical aperture with a 5.0" radius centred on the source.  Integrated intensity maps were created from the CLEAN cubes, which yielded no detections. 

Next, we applied a Keplerian mask to each of the CLEAN image cubes in order to maximise the S/N in the image plane. We extracted spectra and generated integrated intensity maps from the masked cubes, which again resulted in no detections. We also tried stacking the SO 2$_1$-1$_0$ and 8$_8$-7$_7$ lines in the image plane by adding together the integrated intensity maps and spectra. However, SO remained undetected.

Finally, we applied a matched filter to the visibility data, to maximize the S/N in the \emph{uv}-plane \citep{Loomis_2018}. The matched filter technique utilises a template image cube that samples the \emph{uv}-space in order to obtain a set of template visibilities. These can then be used as a filter, which is cross-correlated with the data in an attempt to detect any weak emission. Matched filtering has previously been used to successfully detect weak spectral line features in HD 163296, TW Hya, and HD~100546, providing an improvement in the S/N of up to $\sim$500\% when compared to alternative techniques \citep{Carney2017, Loomis_2018, Booth2018}. We created template emission profiles for each of the spectral windows in the ACA data by modelling the spectral line emission with the DALI thermo-chemical disk modelling code (see section \ref{subsec:chemical_modelling}). The matched filter was then run for each of the spectral lines individually, which again resulted in non-detections for all lines. We derived 3$\sigma$ upper limits for the disk-integrated flux for each of the non-detected spectral lines, calculated from the elliptically masked integrated intensity maps (see Table \ref{table:aca_observations}).

We also report the serendipitous detection of C$^{18}$O 3-2 in the spectral window centred on SO 2$_1$-1$_0$ at 329.385 GHz. The detection is made at a $\sim 24 \sigma$ confidence level, with a peak flux of $7.2 \pm 0.3$ Jy beam$^{-1}$ km s$^{-1}$, as measured from the integrated intensity map. The disk integrated flux is measured using the procedure outlined above, determined to be 5.41 Jy km s$^{-1}$ using a Keplerian mask, and 8.22 Jy km s$^{-1}$ using an elliptical mask.

\subsection{Complementary data}
\label{subsec:complementary_data}

To complement our CS 7-6 data for HD~100546, we make use of a wide range of archival data. Of particular significance to this study are detections of SO 7$_7$-6$_6$ and SO 7$_8$-6$_7$ first presented in \citep{booth_2022} (ALMA program 2019.1.00193.S, PI A. S. Booth). We note that the maximum recoverable scale of the observations is 10.422" ($\sim 1150$ au), which is much larger than the gas disk \citep{Walsh_2014}, and we therefore do not expect any flux missing on short spacings. Similarly, the maximum recoverable scale of our ACA observations is 21.266" ($\sim 2350$ au), also much larger than the gas-disk. The full list of line fluxes and upper limits used to constrain our model is presented in \ref{sec:appendix_obs_data}.

\subsection{Chemical modelling}
\label{subsec:chemical_modelling}

To investigate the origin of the azimuthal asymmetries in various molecular species in the HD~100546 disk, we ran source specific models using the 2D physical-chemical code DALI \citep{Bruderer2012, Bruderer2013}. The code begins with a parameterised gas and dust density distribution and an input stellar spectrum, then uses Monte Carlo radiative transfer to determine the UV radiation field and dust temperature. This provides an initial guess for the gas temperature, which begins an iterative process in which the gas-grain chemistry is solved time-dependantly. Finally, the raytracing module is used to obtain spectral image cubes, line profiles, and disk-integrated line fluxes.

\subsubsection{Disk parameters}
The disk structure is fully parameterised, with a surface density that follows the standard form of a power law with an exponential taper:
\begin{equation}
    \Sigma_\text{gas} = \Sigma_\text{c} \cdot \bigg(\frac{r}{R_c} \bigg)^{-\gamma} \cdot \exp \bigg[- \bigg(\frac{r}{R_c} \bigg)^{2-\gamma} \bigg]
\end{equation}
where $r$ is the radius, $\gamma$ is the surface density exponent, $\Sigma_\text{c}$ is some critical surface density, and $R_\text{c}$ is some critical radius, such that the surface density at $R_\text{c}$ is $\Sigma_\text{c}/e$. The scale height is then given by:
\begin{equation}
    h(r) = h_c\bigg(\frac{r}{R_c}\bigg)^\psi
\end{equation}
where $h_\text{c}$ is the scale height at $R_\text{c}$, and the power law index of the scale height, $\psi$, describes the flaring structure of the disk.

\sigmagas \ and \sigmadust \ extend from the dust sublimation radius (\rsub) \ to the edge of the disk ($R_\text{out}$), and can be varied independantly inside the cavity radius \rcav \ with the multiplication factors \deltagas \ and \deltadust.

The gas-to-dust ratio is denoted \gasdust. Dust settling is implemented by considering two different populations of grains; small grains (0.005-1$\mu$m) and large grains (0.005-1mm). The vertical density structure of the dust is such that large grains are settled towards the midplane, prescribed by the settling parameter $\chi$:

\begin{equation}
    \rho_\text{dust, large} = \frac{f \Sigma_\text{dust}}{\sqrt{2\pi}r \chi h} \cdot \exp \bigg[ -\frac{1}{2} \bigg( \frac{\pi/2 - \theta}{ \chi h} \bigg)^{2} \bigg]
\end{equation}

\begin{equation}
    \rho_\text{dust, small} = \frac{(1-f)\Sigma_\text{dust}}{\sqrt{2\pi}rh} \cdot \exp \bigg[ -\frac{1}{2} \bigg( \frac{\pi/2 - \theta}{h} \bigg)^{2} \bigg]
\end{equation}

where $f$ is the mass fraction of large grains and $\theta$ is the opening angle from the midplane as viewed from the central star. The physical disk parameters used in our model are given in Table \ref{table:modelparameters}.

\subsubsection{Stellar parameters}
HD~100546 is a well-studied 2.49 $\pm 0.02$ \msun \ Herbig Be star of spectral type B9V, with an estimated age of $\sim$ 5 Myr \citep{Arun_2019}. The star is noted for its proximity, located at a distance of 110$\pm$1 pc \citep{GAIACollaborationetal2018}. We note that this differs non-trivially from previous estimates of 97$\pm$4 from \textit{Hipparcos}. The stellar spectrum was modelled by Bruderer et al. (2012) using dereddened FUSE and IUE observations at UV wavelengths, and then extended to longer wavelengths using the B9V template of \citep{Pickles_1998}. The stellar luminosity is $36 \; L_\odot$ \citep{Kama2016b}.

\subsubsection{Chemical network}
The chemical network used in our model is based on a subset of the UMIST 06 \citep{woodall2007} network. It consists of 122 species (including neutral and charged PAHs) and 1701 individual reactions. The code includes H$_2$ formation on dust, freeze-out, thermal and non-thermal desorption, hydrogenation, gas-phase reactions, photodissociation and -ionization, X-ray induced processes, cosmic-ray induced reactions, PAH/small grain charge exchange/hydrogenation, and reactions with vibrationally excited H$_2$. For grain-surface chemistry, only hydrogenation of simple species is considered (C, CH, CH$_2$, CH$_3$, N, NH, NH$_2$, O, and OH). The details of these processes are described more fully in \citep{Bruderer2012}. Of particular relevance to this study, the network includes reactions for 30 sulfur-bearing species, including all those listed in Table \ref{table:aca_observations}. Model parameters of relevance to the disk chemistry are listed in Table \ref{table:modelparameters}.

\subsubsection{Basic fitting procedure}
Our fitting process follows the precedure outlined in \citep{Kama2016b}, making use of additional observational constraints and a larger grid of models. We begin by fitting the SED using a grid of 1728 models, in which the parameters \rgap, $\psi$, $h_c$, \deltadust\ and \gasdust\ are varied. At this stage, \sigmagas\ is kept fixed at an arbitrary value such that changes to \gasdust \ are equivalent to changes only in the dust mass, thus providing us with a baseline estimate for the total dust mass. We find a best fit total dust mass of $1.12 \times 10^{-3}$ \msun, consistent with previous studies \citep{Kama2016b}.

Next, we use the upper limits of the HD 56 $\mu$m and 112 $\mu$m lines to constrain the maximum gas mass. A second grid of models is run in which \sigmagas \ and \gasdust \ are varied in lockstep, allowing the gas mass to vary whilst maintaining the best-fit dust mass. We constrain the total gas mass to $< 5.6 \times 10^{-1}$ \msun, equivalent to a gas-to-dust ratio of \gasdust $\approx 390$ (taking into account dust depletion in the inner cavity). This is not a tight enough constraint to uniquely determine the gas-to-dust ratio, so from this point on we adopt the interstellar value of \gasdust = 100, equivalent to a total gas mass of $1.45 \times 10^{-1}$ \msun. Our model allows for varitions in \gasdust \; within the inner dust cavity, but does not take into account other radial variations such as the observed dust gap in the outer disk between $r \sim 40 - 150$ au.

[C]/[H]$_\text{gas}$ and [O]/[H]$_\text{gas}$ are constrained by modelling the CO ladder, the line profiles of the CO 3-2, CO 6-5 and [CI] transitions, and the radial profile of the CO 3-2 emission from \citep{Walsh_2014}. We find a best fit oxygen abundance of \ohgas\ $\approx (1-7)\times 10^{-5}$. When \ohgas\ > $7 \times 10^{-5}$, the [CI] line is underpredicted or the CO ladder is overpredicted, depending on \chgas. When \ohgas\ < $1 \times 10^{-5}$, the CO ladder is underpredicted for all values of \chgas. Adopting a fiducial oxygen abundance of \ohgas $=2\times 10^{-5}$, we find \chgas$\approx (1-2) \times 10^{-5}$. C$_2$H upper limits presented in \citep{Kama2016b} constrain the global C/O $\lesssim 1$, and we adopt \chgas $=1 \times 10^{-5}$ for our fiducial model, giving a C/O ratio of 0.5. These values are consistent with those found in previous studies of HD~100546 \citep[e.g.][]{Kama2016b}. As with that study, the main uncertainties are due to the limited constraints on the gas-to-dust ratio.

Finally, we constrain the gas-phase elemental sulfur abundance using the disk-integrated CS 7-6, SO 7$_7$-6$_6$ and SO 7$_8$-6$_7$ fluxes and radial intensity profiles, and upper limits for other sulfur-bearing species listed in Table \ref{table:aca_observations}. For this study, we adopt a radially varying sulfur abundance profile, in which [S/H]$=10^{-9}$ between 13-30 au and [S/H]$=10^{-8}$ between 150-230 au, coincident with prominent millimeter dust rings. Outside of these regions, the sulfur is further heavily depleted by a factor of 1000. This is based on high resolution SO observations which suggest some level of correlation between SO emission and dust ring location \citep{booth_2022}. If a single sulfur abundance is used for the entire disk, our model overpredicts SO emission from the outer dust cavity. A detailed study into the volatile sulfur abundance in HD~100546 is the focus of an upcoming companion paper (in prep).

\subsubsection{Modelling azimuthal asymmetries}

We hypothesize that the asymmetries in the observed CS and SO emission can be explained by significant azimuthal variations in the elemental carbon and/or oxygen abundances in HD~100546. In such a scenario, the resulting chemistry leads to an azimuthal disparity in the production of carbon- and oxygen-bearing species.

Our aim is to produce a model in which oxygen-based chemistry dominates in one region of the disk, while carbon-based chemistry dominates in another region of the disk. We investigate whether such a model, using time-dependant chemistry, can self-consistently reproduce the asymmetry in the CS and SO emission.

We simplify our assumptions about the chemistry by constructing a model in which the disk consists of two distinct spatial regions. DALI is a 2-dimensional code which relies upon azimuthal symmetry to produce 3-dimensional outputs, so in order to simulate azimuthal asymmetries it is necessary to run two separate models and splice together the outputs using the following procedure. 

Our starting point is the full-disk model outlined in the previous section, where C/O=0.5 (`model A'). Using the raytracing module in DALI, we obtain spectral image cubes for each transition, with the velocity resolution set to match the observations. Next, we run a second full-disk model in which the carbon and/or oxygen abundances are varied, such that the resulting C/O ratio is different from the first model (`model B'). Using a custom Python script, for each transition the spectral cube from model B is spliced together with its counterpart from model A, following the geometry outlined in Figure \ref{fig_disk_schematic}.  The resulting cube consists of a large `crescent' region taken from model A, in which the C/O ratio is 0.5, and a smaller `wedge' region extracted from model B, with a different C/O ratio. The size of the wedge is prescribed by the angle $\theta$, and the orientation by the angle $\phi$ (measured from the centre of the wedge arc). Angles are measured in the plane of the cube i.e. they are not projected on to the disk, and as such do not directly correlate to angles measured in the disk plane. The result is that the angular region from which the wedge is extracted cuts through very slightly different azimuthal regions for different vertical positions within the disk. A more sophisticated model could take the vertical disk structure into account, but we expect the overall effect on the intensity maps and line profiles to be negligible. We note that that there is precedent in the literature for this kind of chemical modelling \citep{cleeves_2015}. We process the model cubes using the CASA tasks \emph{simobserve} and \emph{simanalyze} in order to create simulated ALMA observations, using parameters that match the observations. Spectra are extracted and the cubes are then collapsed to generate moment 0 integrated intensity maps.

We explored a wide parameter space, considering large scale variations in the carbon/oxygen abundance, wedge size and position, and chemical timescale. Our model grid covers elemental carbon and oxygen abundances that range from $1.0 \times 10^{-6}$ to $2 \times 10^{-4}$, and C/O between 0.3 - 3.0 ($\sim$ 570 models in total). The angular size and position from which we extracted the wedge used parameters that varied from $\theta = 10 - 90^\text{o}$ and $\phi = 0 - 180^\text{o}$. Both the carbon/oxygen abundances and overall size of the wedge are largely constrained by the best-fit full disk model outlined in the previous section; the vast majority of the total disk structure must conform to this model if the general fit is to be maintained. Thus, the wedge region cannot be too large, nor can the carbon/oxygen abundances vary too dramatically without significantly affecting the overall model fit. While large wedge angles better produce the spatial morphology of the SO emission, they lead to disk-integrated CS fluxes that are too high. Smaller wedge angles better reproduce the CS flux, but fail to reproduce the offset in the CS emission from the host star, and lead to SO emission that extends too far around the western side of the disk. 

We simulate the effects of disk shadowing by using the output abundances from the C/O = 0.5 model as input abundances to the chemical network for the high-C/O wedge model, such that the chemical conditions at the beginning of shadowing are similar to that of the unshadowed region of the disk. Before running the chemical network, we decrease the input H$_2$O, CO, and atomic O abundances by varying amounts in order to investigate a range of C/O ratios. The physical justification for depleting each of these species is discussed in \ref{sec:appendix_timescales}. Models are run for a range of chemical timescales, in order to assess how quicky the chemistry rebalances. This method allows us to constrain the CO depletion factor to $\lesssim 0.8$, via comparison with the CO 6-5, CO 3-2, and [CI] 1-0 lines obtained by APEX and previously presented in \citep{Kama2016b} (see \ref{sec:appendix_co_ci}) (the depletion factor is defined as the ratio of the new abundance to the initial abundance). The CO 6-5 line profile has a significant asymmetry in which the blue-shifted peak is $\sim 1.25$x brighter than the red-shifted peak. CO depletion factors higher than $\sim 0.8$ fail to reproduce any significant asymmetry. We test the effects of modelling a range of CO depletion factors between 0 to 0.8, while at the same time varying the level of atomic oxygen and H$_2$O depletion, as to cover a range of C/O ratios. In order to investigate C/O ratios significantly greater than unity, it is necessary to redistribute some of the carbon from the removed CO into other gas-phase species. In this case, we place the carbon into neutral atomic carbon, following \citep{Bergin_2016}. The model presented here uses a CO depletion factor of 0.3, an atomic oxygen depletion factor of 0.3, and an H$_2$O depletion factor of 0 (i.e. total removal), resulting in a gas-phase C/O ratio of 1.5. The model is run to a chemical age of 5 years, based on the approximate shadow transit time (\ref{sec:appendix_timescales}). The full set of parameters used for the model presented in this study are listed in Table \ref{table:modelparameters} (\ref{sec:appendix_modelparams}). Using these values, we find that a wedge size of $\theta = 60^\text{o}$ centred ten degrees north of west west ($\phi = 10^\text{o}$) accurately reproduces both the CS and SO emission morphology and line profiles.

\section*{Data availability}
\label{sec:data_availability}
The data presented here is from the ALMA Cycle 4 program 2016.1.01339.S (PI M. Kama). The raw data is publicly available from the ALMA archive. The reduced data and final imaging products are available upon reasonable request from the corresponding author.

\section*{Corresponding author}
\label{sec:corresponding author}

Correspondence and requests for materials should be addressed to Luke Keyte (luke.keyte.18@ucl.ac.uk).


\begin{acknowledgement}
We acknowledge D.Fedele for sharing the ALMA 870$\mu$m continuum data. L.K. acknowledges funding via a Science and Technology Facilities Council (STFC) studentship. E.F.v.D. is supported by A-ERC grant agreement No. 101019751 MOLDISK. M.N.D. acknowledges the Swiss National Science Foundation (SNSF) Ambizione grant no. 180079, the Center for Space and Habitability (CSH) Fellowship, and the IAU Gruber Foundation Fellowship. C.W. acknowledges financial support from the University of Leeds, the Science and Technology Facilities Council, and UK Research and Innovation (grant numbers ST/T000287/1 and MR/T040726/1).
\end{acknowledgement}

\section*{Author contributions}
LK reduced the ACA CS data, ran the chemical models, performed analysis of both the data and models, and wrote the manuscript. MK contributed to the analysis of both the data and models, original research concepts, and writing of the manuscript, and led the proposal for the ACA data. A.S.B provided the ALMA SO data and contributed to the writing of the manuscript. E.A.B., L.I.C., E.F.v.D., M.N.D., K.F., J.R., O.S., and C.W. contributed to the writing of the manuscript.

\appendix

\clearpage


\section{Assessing the significance of the CS 7-6 emission offset}
\label{sec:appendix_continuum_c18o}

Here, we assess whether the observed offset in the peak of the CS 7-6 emission is related to a physical phenomenon, or the result of an observational error. We begin by comparing the CS emission to the continuum and C$^{18}$O 3-2 observations from our ACA dataset (Figure \ref{fig_offset}). Both the continuum and C$^{18}$O 3-2 line are detected at high signal-to-noise (see Section \ref{sec:methods}). A previous study has suggested that both the continuum and C$^{18}$O 2-1 emission may be slightly asymmetric, with emission peaks that are offset from the star by $\sim 0.08$" and $\sim 0.11$" respectively \citep{miley_2019}. However, such deviations are significantly smaller than the observed CS offset of $\sim1$" (by factors of $\sim 12.5$ and $\sim 9$). At the scale of our observations (using a $\sim 5$" synthesized beam and pixel size of 0.5"), any offset in the continuum or C$^{18}$O emission will be contained entirely within the central pixel. It is therefore reasonable to utilise the continuum and C$^{18}$O emission peaks as reference locations for the host star. We also note that the C$^{18}$O 3-2 spectrum presented here (Figure \ref{fig_offset}, bottom panel) is symmetric about the source velocity, in contrast to the C$^{18}$O 2-1 line presented in \citep{miley_2019}. This suggests that the layer traced by C$^{18}$O 3-2 emission is more symmetric that that traced by the 2-1 emission, and strengthens the case to use it as a frame of reference for the host star. By comparision, it is clear that the CS 7-6 emission is significantly offset from the central star, and the offset cannot therefore be related to the absolute astronometric precision of the observation. We created a residual map by subtracting the CS 7-6 integrated intensity map from the C$^{18}$O 3-2 integrated intensity map. A 1$\sigma$ clip was first applied to the CS data, and the C$^{18}$O peak flux was then scaled to match that of the CS. The residuals were then divided by the rms of the CS data.

Next, we calculate the uncertainty in the beam centroid, given by the following equation \citep{condon_1998}:

\begin{equation}
    \begin{split}
        \sigma_P \approx \frac{1}{2} \sigma \theta / S_P
    \end{split}
    \label{eq:condon98_eq1_1}
\end{equation}

where $\sigma_P$ is the positional uncertainty, $\sigma$ is the rms, $\theta$ is the beam FWHM, and $S_P$ is the peak flux density. Using our reported values of $\sigma = 0.1$ Jy beam$^{-1}$, $\theta = 4.78$", and $S_P = 0.9$ Jy beam$^{-1}$, we find:

\begin{equation}
    \begin{split}
        \sigma_P \approx \frac{1}{2} \times 0.1 \times 4.78 / 0.9 \approx 0.27"
    \end{split}
    \label{eq:condon98_eq1_2}
\end{equation}

We therefore find the CS 7-6 emission peak to be offset by $\sim 1 \pm 0.27$". Since the offset is a factor of $\sim 4$ times the uncertainty, we can be confident that it is at least partly related to a physical characteristic of the source.

Next, we assess the offset in the visibility plane, using the CASA task \emph{uvmodelfit} to fit a model directly to the UV-data. We first split out the spectral windows containing the CS 7-6 emission from continuum-subtracted measurement set, using only the channels centered $\pm 10$ km s$^{-1}$ from the source velocity. We then fit a Gaussian to the data, where the initial guess for shape is based on the synthesized beam properties, and the initial guess for the emission peak is the image centre (ie. location of the star). We run 20 iterations, with the model converging after $\sim 8$. The offset in the emission peak is found to be $0.88 \pm 0.23$" west and $0.68 \pm 0.25$" south, corresponding to a total absolute offset of $1.11 \pm 0.33$". Since \emph{uvmodelfit} can be sensitive to the initial conditions, we repeat the fitting using a range of initial guesses for the location of the emission peak. In all cases the models converge to close to the same result, deviating by a maximum of $\sim8$ \% from the value reported here.

Finally, we looked at data in the ALMA archive for other spectral line observations which might display similar asymmetries. We uncovered an unpublished observation of C$_2$H at 349 GHz (ALMA project 2017.1.00845.S, PI: E. Bergin), obtained at high angular resolution (beam size $\sim 0.3$", approximately 30 au). C$_2$H is a powerful tracer of the C/O ratio, with numerous studies having linked strong C$_2$H emission to chemical environments where C/O > 1 \citep[e.g.][]{Bergin_2016, miotello_2019, bergner_2019, maps_7_bosman2021}. An emission map created from the pipeline data products reveals that C$_2$H is concentrated in a ring at $\sim 230$ au, which approximately coincides with the outer edge of a millimeter dust ring. The total flux on the red-shifted side of the disk is a factor of $\sim 1.5$ greater that that on the shifted side, which may be indicative of an elevated C/O ratio. This is in agreement with our CS 7-6 observation, where the emission is significantly offset towards the red shifted side of the disk. We do not formally present the C$_2$H observatiom here, since a full analysis is well beyond the scope of this work. However, we note that the pipeline data products are freely available in the ALMA archive.

\section{Abundance maps and contribution functions}
\label{sec:appendix_cbfs}

Figure \ref{fig_cbfs} shows CS and SO abundance maps, for both the C/O=0.5 region and C/O>1 wedge . Contribution functions for the CS 7-6 and SO 7$_7$-6$_6$+7$_8$-6$_7$ transitions are overlaid (25\% and 75\% contours).

\section{Varying the size and position of the high-C/O wedge}
\label{sec:appendix_wedge_variations}

Figure \ref{fig_wedge_variations} shows the effect of varying the size and position of the high-C/O wedge on the modelled spectra. The wedge size is described by the arc subtended by the angle $\theta$. The wedge position is described using the azimuthal angle $\phi$ (where $\phi=0$ points due west, and is measured positively anti-clockwise). A given angle $\phi$ corresponds to the position of the arc centre.

We vary the angular size of the high-C/O wedge between $\theta=20^\circ$ to $180^\circ$, while keeping the centre of the wedge arc fixed at $\phi=0$ (Figure \ref{fig_wedge_variations}, top panel). Asymmetries in the SO spectra are more pronounced for larger wedge angles, with increasing wedge sizes leading to decreasing peak fluxes on the red-shifted side of the spectrum. The CS spectrum remains highly asymmetric about the source velocity, regardless of wedge size, with larger wedges result in broader line profiles and higher peak fluxes.

We vary the angular position of the high-C/O wedge between $\phi = -60^\circ$ to $60^\circ$, while keeping the angular size fixed at $\theta=60^\circ$ (Figure \ref{fig_wedge_variations}, bottom panel). Asymmetries in the SO spectra are most pronounced when the wedge points towards $60^\circ$, since this corresponds to the region of the disk in which line-of-sight velocities are most highly red-shifted. Rotating the arc position clockwise through to $-60^\circ$ results in a incrementally smaller contribution to the SO spectra from the highly red-shifted part of the disk, and therfore a more symmetric double-peaked structure. Similarly, the CS spectrum is most highly red-shifted when the high-C/O wedge is positioned at $60^\circ$, with a narrow profile and high peak flux. As the arc position is rotated clockwise through to $-60^\circ$, the line becomes broader with smaller peak flux, with the peak of the line profile moving closer to the source velocity.

\section{Temperature and density maps}
\label{sec:appendix_tempdens}

Figure \ref{fig_temp_density} shows the temperature and density maps for dust and gas in the HD 100546 disk model.

\section{Cooling, freeze-out, and chemical timescales}
\label{sec:appendix_timescales}

We propose that material associated with a protoplanet in HD 100546's inner dust cavity casts a shadow on a region of the outer disk, resulting in the azimuthal chemical variations described in this work. In order for such material to cast a shadow covering an angular region of $60^\circ$, as prescribed by our model, the azimuthal extent of the material must be much larger than the host star. At a radial distance of 13 au, which is equivalent to the outer edge of the dust cavity, the necessary azimuthal extent of the shadowing material is $\sim 15$ au. Additionally, the material must extend vertically such that the shadow enables freeze-out of H$_2$O in the upper layers of the disk atmosphere. Taking $r=100$ au and $z/r=0.25$ as a representative location for excess H$_2$O freeze-out (via comparison with the modelled H$_2$O snow surface, see Figure \ref{fig_tdust_snowsurface}), the necessary vertical extent of the shadowing material is $\sim 3.2$ au. This is reasonable, since hydrodynamic simulations of disks show that dust concentrations associated with large embedded protoplanets can be highly elongated, especially in the $\phi$ direction \citep[e.g.][]{zhu_2014}.

If the material responsible for shadowing orbits at 13 au, it will complete one orbit in $\sim 30$ years. A shadow covering an azimuthal region of $60^\circ$ will therefore transit a given region of the disk in $30 \times (60/360) = 5$ years. However, since the shadowed material is in orbit itself, the total time spent in shadow will vary as a function of orbital separation. Shadowing timescales are longest in the inner disk, where the orbital period of the material being shadowed is comparible to the orbital period of the shadow. This falls off rapidly with radius, approaching the minimum shadowing time of $\sim5$ years in the outer disk. Since reproducing the offset in the CS emission requires a large increase in the CS abundance in the outer disk, we take 5 years as our upper limit on the time available for changes in the chemistry to occur such that the C/O ratio becomes super-solar. To assess the feasibility of the shadowing scenario, we present a simple analytical model that takes into consideration three important timescales; cooling, freeze-out, and chemical conversion.

The cooling timescale for a single dust grain is given by:

\begin{equation}
    \begin{split}
        t_\text{cooling} &= \frac{mC\Delta T}{Q_\text{rad}P}
    \end{split}
    \label{eq:cooling_timescale}
\end{equation}

where $m$ is the mass of the dust grain, $C$ is the specific heat capacity, $\Delta T$ is the change in temperature, and $P=A\sigma T^4$ is the radiative power of the dust grain (where A is the grain surface area, and $\sigma$ is the Stefan-Boltzmann constant). $Q_\text{rad}$ is the Planck mean efficiency for dust grains at temperatures between 10-250 K (where $Q_\text{rad} = 1.25 \times 10^{-5} T_d^2 (r/1\mu m)$, where $r$ is the grain radius \citep{tielens_book}).

We make the simplifying assumption that gas in the disk atmosphere is coupled to a population of small dust grains of equal size $r=1\mu m$. Taking $\rho = 4$ g cm$^{-3}$ as a typical value for grain density (based on silicate materials such as olivine), the grain mass is then $m=\rho(4/3 \pi r^3) \approx 1.68 \times 10^{-14}$ kg. We take $C=2000$ J kg$^{-1}$ K$^{-1}$, which is again typical of silicate material. Finally, we make the simplification that all H$_2$O is in gaseous form at 100 K and in ice at 50 K i.e. $\Delta T = 50$ K. Note that these approximate values are lower than the canonical freezeout temperature for H$_2$O to account for the effects of photodesorption. For grains with radius $r=1 \mu m$ at 100 K, $Q_\text{rad} = 0.125$. The cooling timescale can then be calculated as:

\begin{equation}
    \begin{split}
        t_\text{cooling} &= \frac{1.68 \times 10^{-14} \times 2000 \times (100-50)}{0.125 \times 4\pi (10^{-6})^2 \times 5.67 \times 10^{-8} \times 100^4} \\
          &\approx 189 \text{ s}
    \end{split}
    \label{eq:cooling_timescale2}
\end{equation}

We now consider the freeze-out timescale, which is dictated by the efficiency with which gas-phase H$_2$O molecules collide with cooled grains. The mean free path of a gas molecule is given by:

\begin{equation}
    \mu = \frac{1}{n\pi r^2}
    \label{eq:mean_free_path}
\end{equation}

where $n$ is the number density of the dust grains, and $r$ is the grain radius. We extract values for $n$ directly from our chemical model, from a range of grid cells that trace the location of the H$_2$O freeze-out in the upper disk atmosphere, then reduce the value by a factor of 100 to account for dust settling to the midplane.

The average gas speed of a molecule is given by:

\begin{equation}
    v_\text{rms} = \sqrt{\bar{v}^2} = \sqrt{\frac{3kT}{m}}
    \label{eq:gas_speed}
\end{equation}

where $k$ is the Boltzmann constant and $m$ is the mass of the H$_2$O molecule. The freeze-out timescale, which is equal to the collision timescale, is then given by:

\begin{equation}
    \begin{split}
        t_\text{freeze} &= \frac{\mu}{v_\text{rms}} \\
    \end{split}
    \label{eq:freeze-out}
\end{equation}

Note that it is not necessary for the bulk of the gas to cool down to the dust temperature for freezeout to occur when the gas and dust are thermally decoupled, as is the case in the disk atmosphere (Extended Data Fig. 3). The increase in dust temperature due to a collision will be extremely small, and since the dust cooling timescale is extremely short this can be considered negligible. The timescale for the dust and gas to reach thermal equilibrium will be longer that the timescale derived here, but our calculation is reasonable when considering freezeout in the conditions described above.

We compute $t_\text{freeze}$ for a range of gas-to-dust ratios between $\Delta_\text{g/d} = 10 - 390$, where the lower limit is taken from \citep{Kama2016b} and the upper limit is based on constraints from modelling HD lines (see Section \ref{subsec:chemical_modelling}). We find that $t_\text{freeze}$ can vary from $\sim 47$ hours to $4500$ years between $r=13$ to $300$ au, depending on $\Delta_\text{g/d}$. These values are subject to large uncertainties because $n_\text{dust}$ can vary quite dramatically depending on the precise location of the disk atmosphere from which it is extracted, in addition to uncertainties in $\Delta_\text{g/d}$ due to dust settling and radial drift. While the simple calculations we present here relate to a single grain, these can reasonably be used as an approximation for the entire local grain population in the optically thin regime, which is a reasonable assumption for material in the upper surface layers of the disk.

Finally, we consider the relevant chemical timescales. We simulate the effects of shadowing by first extracting the final elemental and molecular abundances from our `unshadowed' C/O=0.5 model, using these as an input for the shadow model. The total elemental gas-phase oxygen is then depleted by setting the input H$_2$O abundance to zero (representing total freeze-out). This complete removal of H$_2$O is not enough to increase the gas-phase C/O above unity, since the majority of element oxygen is tied up in atomic O and CO. However, our chemical models show that when H$_2$O is removed from the gas phase on such a large scale, the following chemical rebalance leads to a significant production of new gas-phase H$_2$O via the following pathway:

\begin{gather}
    \text{H}_2 \text{ + O} \rightarrow \text{OH + H}
    \label{eq:chem1}\\
    \text{H}_2 \text{ + OH} \rightarrow \text{H}_2 \text{O + H}
    \label{eq:chem2}
\end{gather}

The subsequent freeze-out of this newly-produced gas-phase H$_2$O therefore provides a route by with atomic oxygen can be removed from the gas phase, further elevating the C/O ratio. Additional oxygen (and carbon) may be removed by the conversion of gas-phase CO into CO$_2$ ice, via a surface reaction with OH radicals produced by UV irradiation of amorphous solid water:

\begin{gather}
    \text{CO + OH} \rightarrow trans-\text{HOCO}
    \label{eq:co2_1}\\
    trans-\text{HOCO} \rightarrow cis-\text{HOCO}
    \label{eq:co2_2}\\
    cis-\text{HOCO} \rightarrow \text{CO}_2 \text{ + H}
    \label{eq:co2_3}
\end{gather}

We can expect reactions \ref{eq:co2_1} - \ref{eq:co2_3} to be enhanced in the shadowed region, since the lower temperature locks a larger fraction of H$_2$O into ices. Laboratory experiments show that this pathway can permanently sequester gas-phase CO into CO$_2$ ice at temperatures between 40-60 K (well above the canonical CO freeze-out temperature), with an efficiency of up to 27\% \citep{van_scheltinga_2022}. While reactions \ref{eq:co2_1} - \ref{eq:co2_3} are not included in our chemical network, we take them into account by manually adjusting the abundances as follows. We adjust the initial conditions of our shadow model, depleting not only H$_2$O, but also a fraction of atomic O and CO, such that the resulting gas-phase C/O > 1.0 (see Section \ref{subsec:chemical_modelling}). We let the chemistry evolve and extract the CS and SO abundances as a function of time from key regions of the disk. We find that the newly elevated C/O ratio favours both the formation of CS and destruction of SO, primarily via the reactions:

\begin{gather}
    \text{C + SO} \rightarrow \text{CS + O}
    \label{eq:chem3}\\
    \text{C + SO} \rightarrow \text{CO + S}
    \label{eq:chem4}
\end{gather}

These reactions have no activation barrier and so can proceed in cold gas. CS is formed at a faster rate than SO is destroyed, since there are prominent CS formation pathways in addition to reaction \ref{eq:chem3}:

\begin{gather}
    \text{CH}_2 \text{ + S} \rightarrow \text{CS + H}_2
    \label{eq:chem5}\\
    \text{C + HS} \rightarrow \text{CS + H}
    \label{eq:chem6}
\end{gather}

The effect of this time evolution on the CS and SO abundances and disk integrated line fluxes is illustrated in Figure \ref{fig_chemistry}. The CS 7-6 disk-integrated line flux steadily increases with time in the super-stellar C/O environment, with the SO 7$_7$-6$_6$ disk-integrated line flux showing the opposite trend. To approximate the timescale at which this phenomenon becomes apparent in observations, we define the chemical timescale as the period in which the CS abundance increases by a factor of two, and the SO abundance decreases by a factor of two, whichever is longest (see Figure \ref{fig_chemistry}). We calculate the chemical timescale at different radial regions in the disk, by extracting the variations in CS and SO abundance from key grid cells in the model (Figure \ref{fig_chemistry}, lower panels). We find that the chemical timescale varies between $\sim 2 \times 10^{-4}$ to $\sim 5$ years, for radial regions between 13-300 au.

The combined results for cooling, freeze-out, and chemical conversion are illustrated in Figure \ref{fig_timescales}. The total time is almost entirely dependent on the freeze-out timescale, since the cooling and chemical timescales are negligable compared to the shadow transit time. We show that, within the $\sim 5$ year period it takes the shadow to transit a given region of the disk, the gas-phase C/O ratio can become super-solar within the inner $\sim 200$ au of the disk (within $\sim 150$ au for $\Delta_\text{g/d}=100$). Considering the fact that prominent emission from both CS and SO emanates from inside of $\sim 200$ au (\ref{sec:appendix_cbfs}), and the uncertainties associated with the freeze-out timescale, we conclude that shadowing is a plausible means by which the local C/O ratio can become elevated, leading to the spatial and spectral asymmetries presented in this work. 

A more sophisticated model could take into account the reverse timescale, and the cyclical nature of disk shadowing, by addressing the effects of many transits over the lifetime of the disk. Such an investigation is beyond the scope of this work.

\section{CS 7-6 channel maps}
\label{sec:appendix_channel_maps}

Figure \ref{fig_channel_maps} shows the individual channel maps for the CS 7-6 detection with the ACA.

\section{CO 6-5, CO 3-2, and [CI] spectra}
\label{sec:appendix_co_ci}

Figure \ref{fig_coci} shows the modelled and observed CO 6-5, CO 3-2, and [CI] 1-0 spectral lines, used to constrain the level of CO depletion in the high-C/O wedge (see Section \ref{subsec:chemical_modelling}).

\section{Disk model parameters}
\label{sec:appendix_modelparams}

The parameters used in our disk models are listed in Table \ref{table:modelparameters}.

\section{Observational data}
\label{sec:appendix_obs_data}

The observational parameters for each of the spectral windows in the ACA dataset are given in Table \ref{table:aca_observations}. The full list of line fluxes and upper limits used to constrain our model is given in Table \ref{table:model_constraints}.

\clearpage


\begin{figure*}[h!]
\centering
\includegraphics[width=\textwidth]{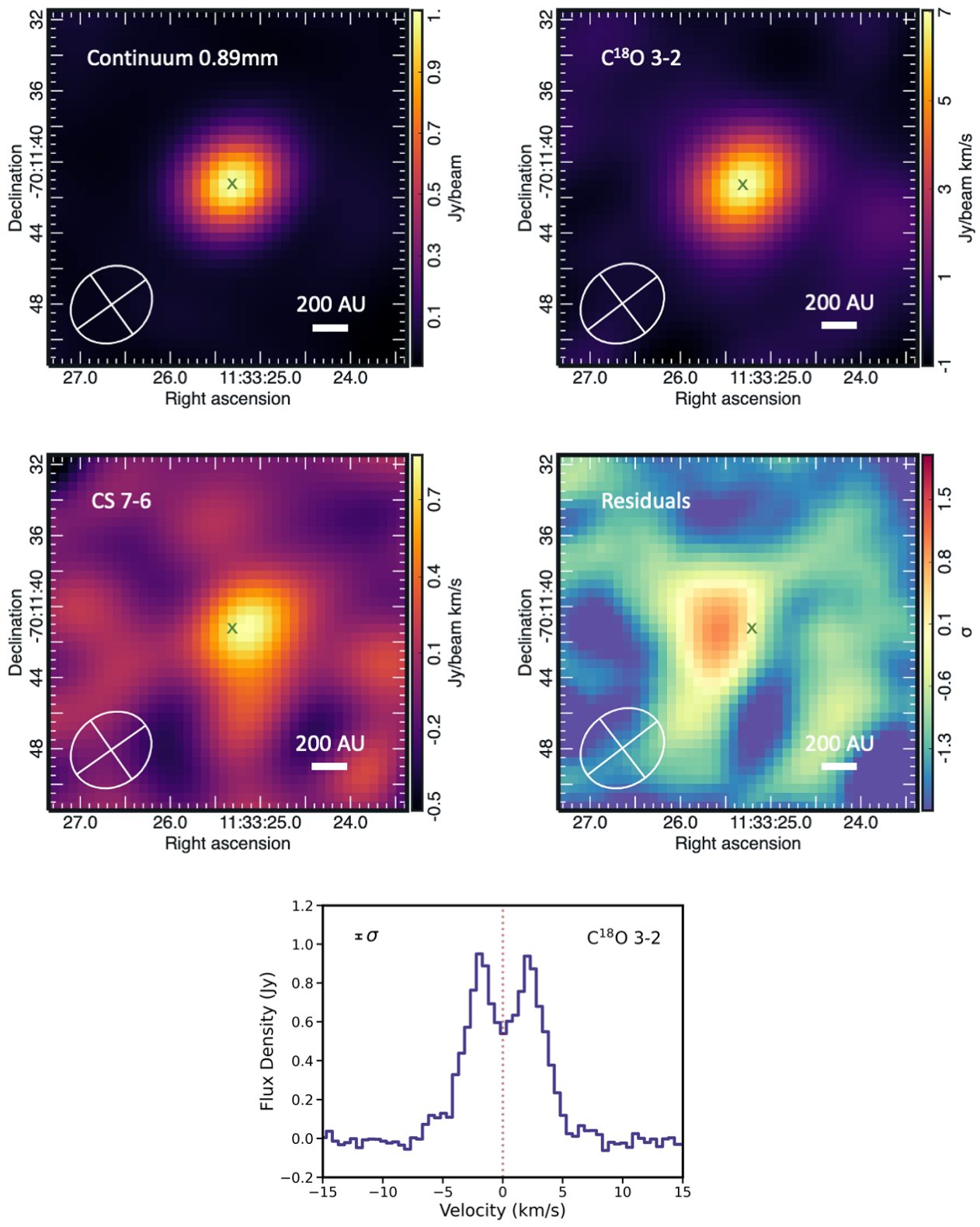}
\caption{\textbf{Comparison between continuum, C$^{18}$O, and CS ACA data}. \emph{Top left: }Continuum emission map. \emph{Top right: } C$^{18}$O 3-2 integrated intensity map. \emph{Middle left: }CS 7-6 integrated intensity map. \emph{Middle right: }Residual map created by subtracting the CS emission from the C$^{18}$O emission. The C$^{18}$O flux was scaled to match the peak flux of the CS emission, and a 1$\sigma$ clip was applied to the CS emission. The residuals are divided by the rms of the CS data, from which it is evident that the CS emission is significantly offset from the star (denoted with `x'). \emph{Bottom: }C$^{18}$O 3-2 spectrum.}
\label{fig_offset}
\end{figure*}

\clearpage

\begin{figure*}[hbt!]
\centering
\includegraphics[width=\textwidth]{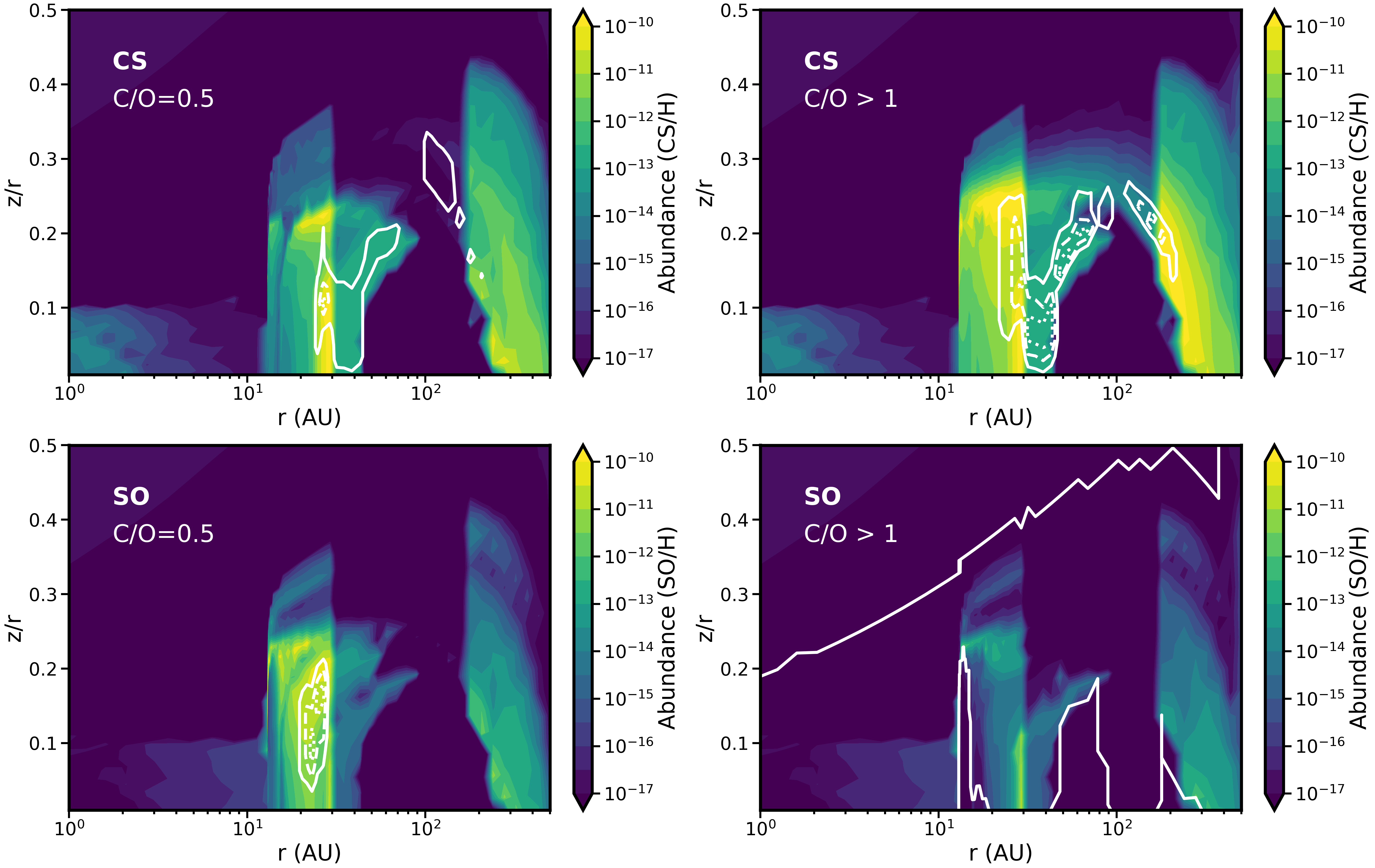}
\caption{\textbf{Abundance maps and contribution functions for the modelled CS and SO emission in HD 100546}. Each panel shows an abundance map overlaid with contours representing 25\% and 75\% line emission (white). Top: CS 7-6 emission from the C/O=0.5 region (\emph{left}) and C/O>1 wedge (\emph{right}). Bottom: SO $7_7-6_6$+$7_8-6_7$ emission from the C/O=0.5 region (\emph{left}) and C/O>1 wedge (\emph{right}).}
\label{fig_cbfs}
\end{figure*}

\clearpage

\begin{figure*}[hbt!]
\centering
\includegraphics[clip=,width=\linewidth]{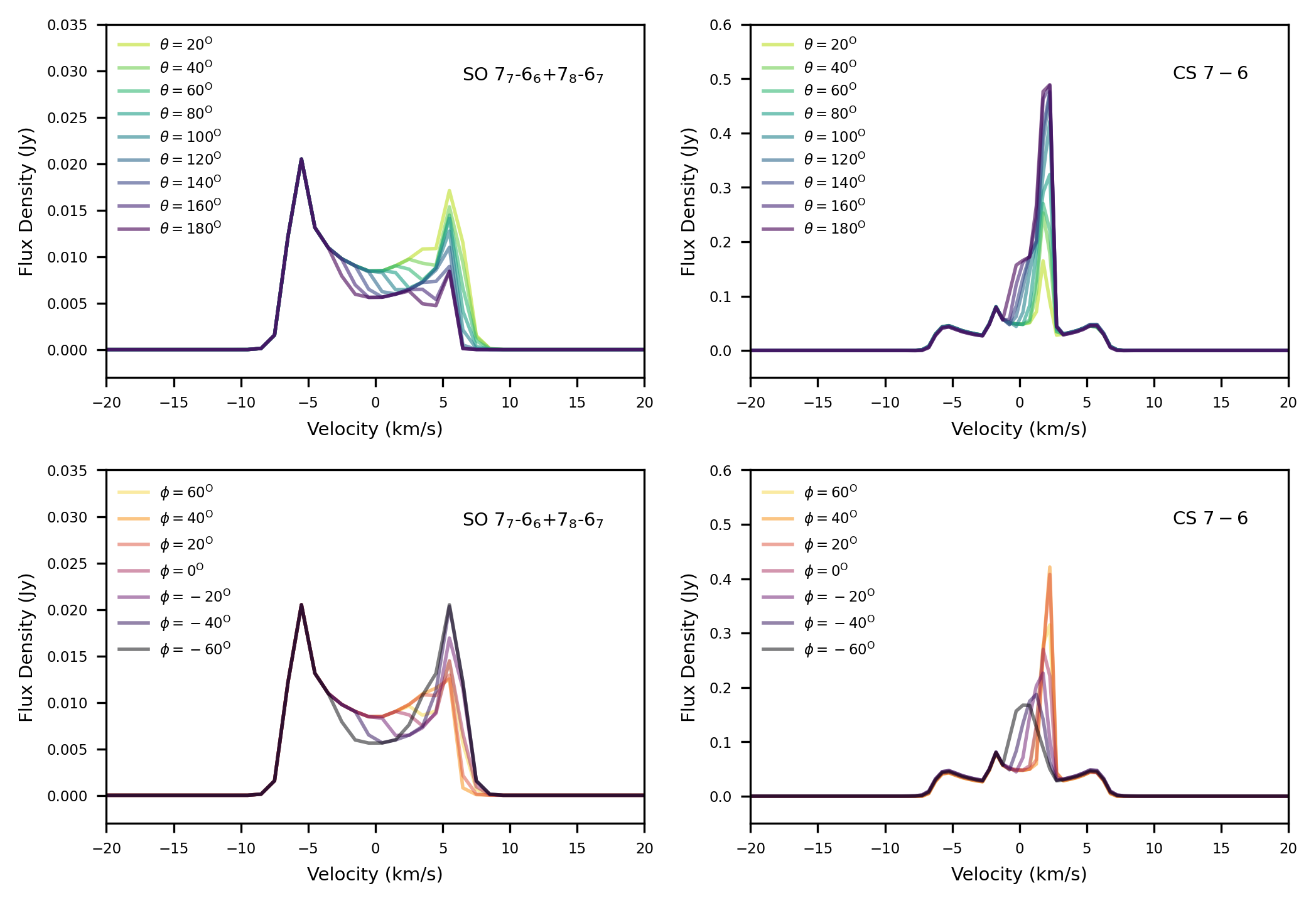}
\caption{\textbf{Effect of varying the high-C/O wedge size and position on the modelled spectra}. \emph{Top panel: }SO 7$_7$-6$_6$+7$_8$-6$_7$ (left) and CS $7-6$ (right) spectra for variations in wedge size ($\theta$), centered on position $\phi=0$. \emph{Bottom panel: }SO 7$_7$-6$_6$+7$_8$-6$_7$ (left) and CS $7-6$ (right) spectra for variations in wedge position ($\phi$), for a fixed angular size $\theta=60^\circ$.}
\label{fig_wedge_variations}
\end{figure*}

\clearpage

\begin{figure*}[hbt!]
\centering
\includegraphics[width=\textwidth]{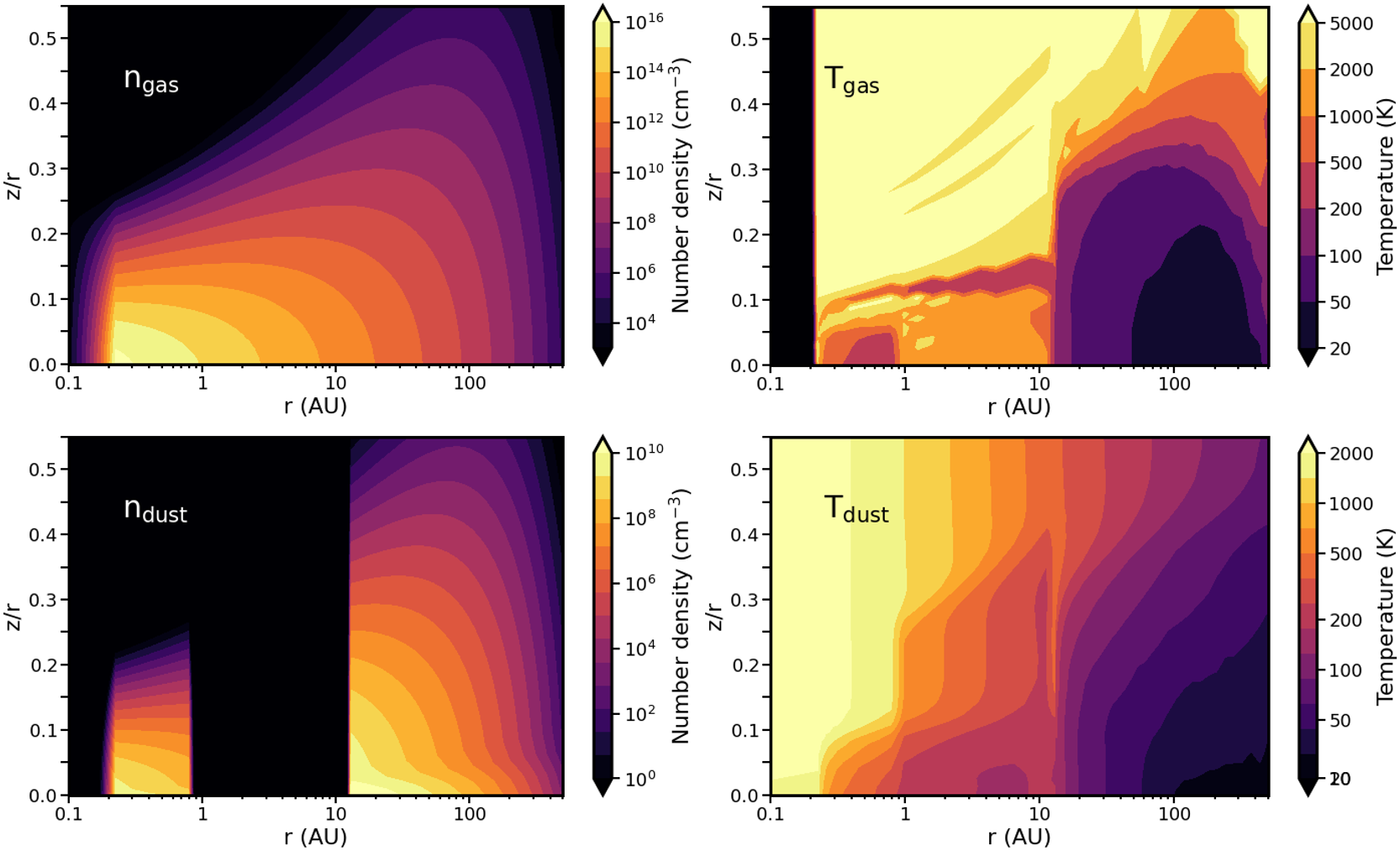}
\caption{\textbf{Temperature and density maps for the baseline HD 100546 disk model (C/O=0.5)}.\emph{ Top left: }Gas number density. \emph{Bottom left: }Dust number density. \emph{Top right: }Gas temperature. \emph{Bottom right: }Dust temperature.}
\label{fig_temp_density}
\end{figure*}

\clearpage

\begin{figure*}[ht!]
\centering
\includegraphics[clip=,width=0.9\linewidth]{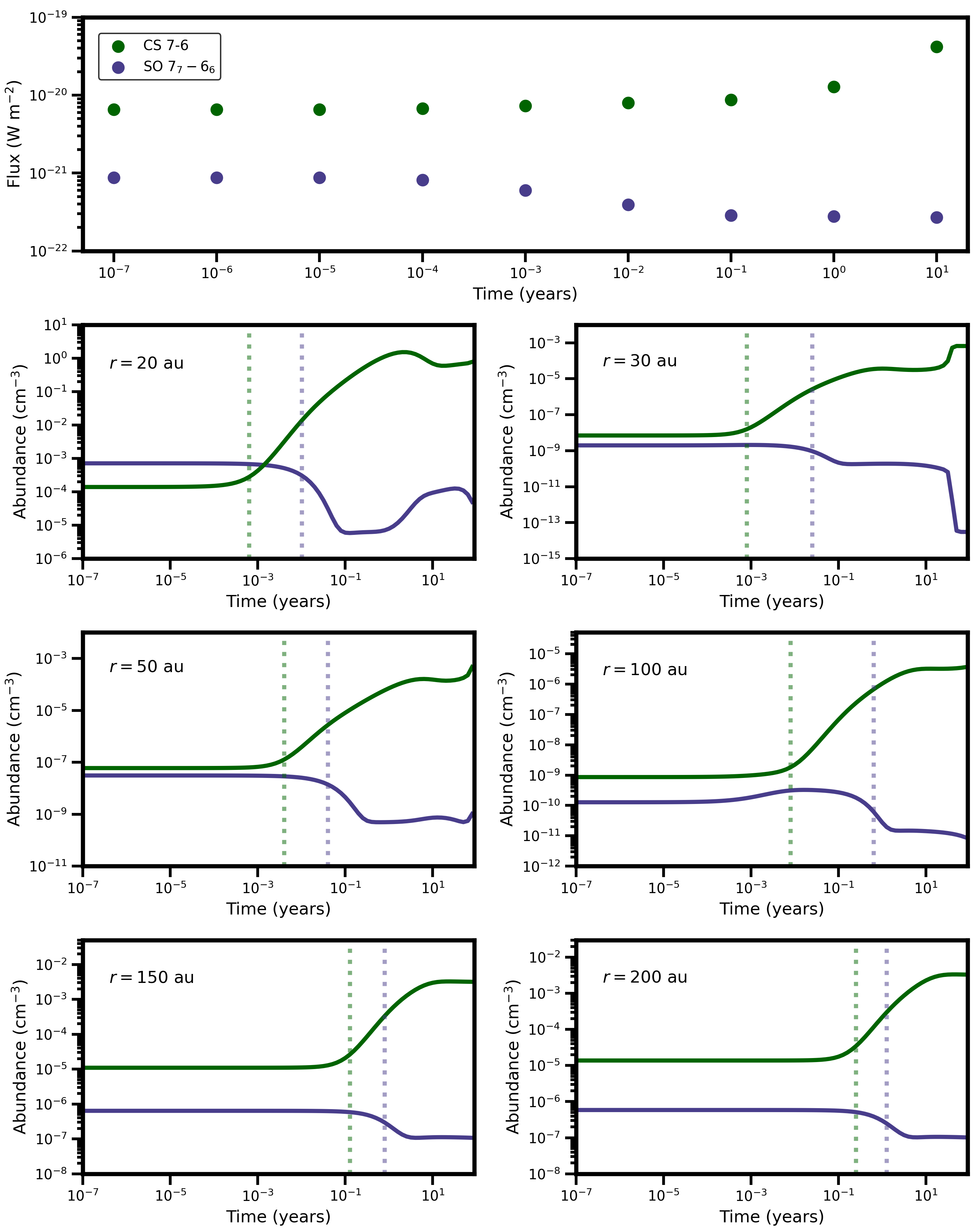}
\caption{\textbf{Time-dependent disk-integrated fluxes and abundances in the high C/O wedge}. \emph{Top panel: }Disk-integrated CS $7-6$ (green) and SO $7_7-6_6$ (purple) fluxes, as a function of time in the region where C/O>1. \emph{All other panels: } CS (green) and SO (purple) abundances at different radial locations, as a function of time in the region where C/O>1. All values are extracted from the model at z/r $\sim 0.25$. Dotted lines represent the point in time where the CS flux increases by a factor of two from its initial values (green), and the SO flux decreases by a factor of two from its initial value (purple). The longer of these two times is taken as the chemical timescale at that particular radius.}
\label{fig_chemistry}
\end{figure*}

\clearpage

\begin{figure*}
\centering
\includegraphics[clip=,width=1\linewidth]{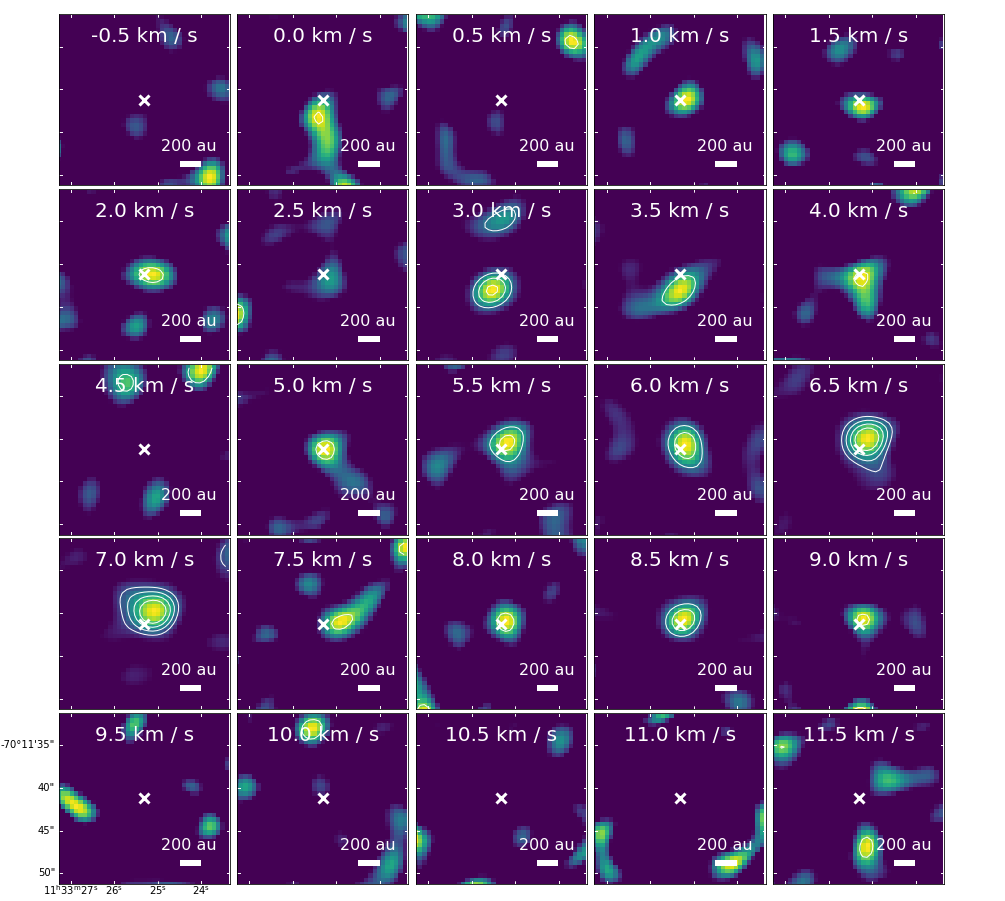}
\caption{\textbf{CS 7-6 channel maps at a spectral resolution of 0.5 km s$^{-1}$}. Contours are at 35\%, 50\%, 65\%, and 80\% peak flux. The source velocity is $V_\text{LSRK} = 5.7 \text{ km s}^{-1}$. The white cross denotes the position of the star.}
\label{fig_channel_maps}
\end{figure*}

\clearpage

\begin{figure*}
\centering
\includegraphics[clip=,width=0.5\linewidth]{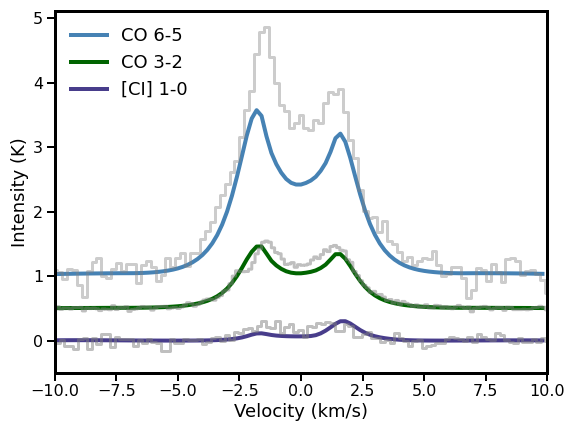}
\caption{\textbf{Modelled and observed CO and [CI] spectra}. Modelled CO 6-5 (blue), CO 3-2 (green), and [CI] 1-0 (purple) spectra. APEX observations in grey. The model reproduces the asymmetry of the CO 6-5 spectral line (but underpredicts the peak CO 6-5 flux, as in previous studies of HD~100546 \citep{Kama2016b}).}
\label{fig_coci}
\end{figure*}

\clearpage

\begin{table*}
\caption{HD~100546 disk model parameters.}             
\label{table:modelparameters}      
\centering
\begin{tabular}{l l l}     %
\hline\hline       
                      
Parameter & Description & Fiducial\\ 
\hline                    
   \rsub                & Sublimation radius                       & 0.25 au                               \\
   \rgap                & Inner disk size                          & 1.0 au                                \\
   \rcav                & Cavity radius                            & 13 au                                 \\
   $R_\text{out}$       & Disk outer radius                        & 1000 au                               \\
   $R_c$                & Critical radius for surface density      & 50 au                                 \\
   \deltagas            & Gas depletion factor inside cavity       & 1                                     \\
   \deltadust           & Dust depletion factor inside cavity      & $10^{-4}$                             \\
   $\gamma$             & Power law index of surface density profile   & 1.0                                   \\
   $\chi$               & Dust settling parameter                      & 0.2                                   \\
   $f$                  & Large-to-small dust mixing parameter         & 0.85                                  \\
   $\Sigma_c$           & $\Sigma_\text{gas}$ at $R_c$                 & 82.75 g cm$^{-2}$                     \\
   $h_c$                & Scale height at $R_c$                        & 0.10                                  \\
   $\psi$               & Power law index of scale height              & 0.20                                  \\
   \gasdust             & Gas-to-dust mass ratio                       & 100                                   \\
   $L_*$                & Stellar luminosity                           & $36\; L_\odot$                        \\
   $L_X$                & Stellar X-ray luminosity                     & $7.94 \times 10^{28} \text{ erg s}^{-1}$    \\
   $T_X$                & X-ray plasma temperature                     & $7.0 \times 10^{7}$ K                 \\
   $\zeta_\text{cr}$    & Cosmic ray ionization rate                   & $3.0 \times 10^{-17}$ s$^{-1}$        \\
   $M_\text{gas}$       & Disk gas mass                                & $1.45 \times 10^{-1}$ \msun           \\
   $M_\text{dust}$      & Disk dust mass                               & $1.12 \times 10^{-3}$ \msun           \\
   $\text{[C/H]}_\text{gas}$        & Initial carbon abundance (relative to hydrogen)  & $1.0 \times 10^{-5}$ (where C/O=0.5)\\
   $\text{[O/H]}_\text{gas}$        & Initial oxygen abundance (relative to hydrogen)   & $2.0 \times 10^{-5}$ (where C/O=0.5)\\
   $t_\text{chem}$      & Timescale for time-dependent chemistry   & 5 Myr (where C/O=0.5) \\
\hline                  
\end{tabular}
\end{table*}

\begin{table*}
\caption{Observational parameters for molecular species in HD 100546 from the ACA (program 2016.1.01339.S). Upper limits are at 3$\sigma$ level, denoted by `<'. Detections are given for Keplerian masked cubes (\emph{a}) and elliptically masked cubes (\emph{b}).}
\label{table:aca_observations}      

\centering

\begin{tabular}{l l l l l l l l l}     
\hline\hline       
                      
Molecule & Transition & $\nu$ (GHz) & $\Delta_{\nu}$ (GHz) & E$_\text{up}$ (K) & A$_\text{ul}$ (s$^{-1}$) & Beam Size & RMS (Jy beam$^{-1}$ km s$^{-1}$) & Flux (Jy km s$^{-1}$)\\
\hline
  SO                & $2_1$-$1_0$              &  $329.385$   & $0.250$ & $15.8$ & $1.423 \times 10^{-5}$   & $4.83" \times 4.08"$ &    $0.14$     &  $< 0.66$   \\
  SO$_2$            & $4_{3,1}$-$3_{2,2}$      &  $332.505$   & $0.250$ & $31.3$ & $3.290 \times 10^{-4}$   & $4.84" \times 4.07"$ &    $0.10$     &  $< 0.48$   \\
  C$^{36}$S         & $7-6$                    &  $332.510$   & $0.250$ & $63.9$ & $6.951 \times 10^{-4}$   & $4.78" \times 4.07"$ &    $0.10$     &  $< 0.46$   \\
  HCS$^+$           & $8-7$                    &  $341.339$   & $0.250$ & $73.7$ & $8.352 \times 10^{-4}$   & $4.62" \times 4.03"$ &    $0.14$     &  $< 0.71$   \\
  CS                & $7-6$                    &  $342.883$   & $0.250$ & $65.8$ & $8.368 \times 10^{-4}$   & $4.78" \times 4.06"$ &    $0.10$     &  $0.62^a$, $1.02^b$      \\
  H$_2$CS           & $10_{0,10}$-$9_{0,9}$    &  $342.946$   & $0.250$ & $90.6$ & $6.080 \times 10^{-4}$   & $4.73" \times 3.90"$ &    $0.31$     &  $< 1.50$    \\
  $^{34}$SO         & $2_3$-$2_1$              &  $343.851$   & $0.250$ & $20.9$ & $1.382 \times 10^{-7}$   & $4.63" \times 3.89"$ &    $0.31$     &  $< 1.60$    \\
  SO                & $8_8$-$7_7$              &  $344.310$   & $0.250$ & $87.5$ & $5.188 \times 10^{-4}$   & $4.62" \times 3.89"$ &    $0.30$     &  $< 1.50$    \\
  C$^{18}$O         & $3-2$                    &  $329.330$   & $0.250$ & $31.6$ & $2.172 \times 10^{-6}$   & $4.96" \times 4.21"$ &    $0.30$     &  $5.41^a$, $8.22^b$\\
\hline                  
\end{tabular}
\end{table*}

\clearpage

 \onecolumn

\begin{table*}
\centering
\caption{Disk-integrated line fluxes and upper limits used to constrain our model.}\label{table:model_constraints}
\scalebox{0.7}{
\begin{tabular}{llll}
\toprule
Molecule          & Transition     & Flux (W m$^{-2}$)               & Reference \\
\midrule 
CO                & 3-2            & $1.72\pm 0.04 \times 10^{-18}$   & \citep{Kama2016b} \\
CO                & 6-5            & $1.61\pm 0.08 \times 10^{-17}$   & \citep{Kama2016b} \\
CO                & 7-6            & $2.35\pm 0.27 \times 10^{-17}$   & \citep{vanderwiel2014} \\
CO                & 8-7            & $3.31\pm 0.35 \times 10^{-17}$   & \citep{vanderwiel2014} \\
CO                & 9-8            & $4.53\pm 0.41 \times 10^{-17}$   & \citep{vanderwiel2014} \\
CO                & 10-9           & $5.53\pm 0.37 \times 10^{-17}$   & \citep{vanderwiel2014} \\
CO                & 11-10          & $5.13\pm 0.45 \times 10^{-17}$   & \citep{vanderwiel2014} \\
CO                & 12-11          & $5.83\pm 0.33 \times 10^{-17}$   & \citep{vanderwiel2014} \\
CO                & 13-12          & $6.07\pm 0.47 \times 10^{-17}$   & \citep{vanderwiel2014} \\
CO                & 14-13          & $6.00\pm 0.83 \times 10^{-17}$   & \citep{meeus2012} \\
CO                & 15-14          & $8.83\pm 0.97 \times 10^{-17}$   & \citep{meeus2012} \\
CO                & 16-15          & $5.88\pm 0.97 \times 10^{-17}$   & \citep{meeus2012} \\
CO                & 17-16          & $7.39\pm 1.0 \times 10^{-17}$    & \citep{meeus2012} \\
CO                & 18-17          & $7.15\pm 0.69 \times 10^{-17}$   & \citep{meeus2012} \\
CO                & 19-18          & $6.47\pm 0.85 \times 10^{-17}$   & \citep{meeus2012} \\
CO                & 20-19          & $4.99\pm 0.57 \times 10^{-17}$   & \citep{meeus2012} \\
CO                & 21-20          & $6.50\pm 0.88 \times 10^{-17}$   & \citep{meeus2012} \\
CO                & 22-21          & $\leq4.2 \times 10^{-17}$        & \citep{meeus2012} \\
CO                & 23-22          & $7.84\pm 1.1 \times 10^{-17}$    & \citep{meeus2012} \\
CO                & 24-23          & $7.13\pm 1.2 \times 10^{-17}$    & \citep{meeus2012} \\
CO                & 25-24          & $\leq7.71 \times 10^{-17}$       & \citep{meeus2012} \\
CO                & 28-27          & $8.15\pm 1.1 \times 10^{-17}$    & \citep{meeus2012} \\
CO                & 29-28          & $7.86\pm 1.9 \times 10^{-17}$    & \citep{meeus2012} \\
CO                & 30-29          & $7.34\pm 1.5 \times 10^{-17}$    & \citep{meeus2012} \\
CO                & 31-30          & $\leq1.40 \times 10^{-16}$       & \citep{meeus2012} \\
CO                & 32-31          & $\leq6.53 \times 10^{-17}$       & \citep{meeus2012} \\
CO                & 33-32          & $\leq8.45 \times 10^{-17}$       & \citep{meeus2012} \\
CO                & 34-33          & $5.31\pm 1.4 \times 10^{-17}$    & \citep{meeus2012} \\
CO                & 35-34          & $\leq4.41 \times 10^{-17}$       & \citep{meeus2012} \\
CO                & 36-35          & $5.29\pm 1.3 \times 10^{-17}$    & \citep{meeus2012} \\
CO                & 37-36          & $\leq8.29 \times 10^{-17}$       & \citep{meeus2012} \\
CO                & 38-37          & $\leq1.07 \times 10^{-16}$       & \citep{meeus2012} \\
$^{13}$CO         & 3-2            & $\leq 6.6 \times 10^{-19}$       & \citep{panic2010} \\
$^{13}$CO         & 6-5            & $\leq 7.5 \times 10^{-18}$       & \citep{vanderwiel2014} \\
$^{13}$CO         & 7-6            & $\leq 7.2 \times 10^{-18}$       & \citep{vanderwiel2014} \\
$^{13}$CO         & 8-7            & $\leq 1.02 \times 10^{-17}$      & \citep{vanderwiel2014} \\
$^{13}$CO         & 9-8            & $\leq 1.44 \times 10^{-17}$      & \citep{vanderwiel2014} \\
$^{13}$CO         & 11-10          & $\leq 1.68 \times 10^{-17}$      & \citep{vanderwiel2014} \\
OI                & 145$\mu$m      & $3.57\pm 0.13 \times 10^{-16}$   & \citep{meeus2012, fedele2013} \\
\toprule
\end{tabular}

\hspace{2em}

\begin{tabular}{llll}
\toprule
Molecule          & Transition     & Flux (W m$^{-2}$)               & Reference \\
\midrule 
OI                & 63$\mu$m       & $5.54\pm 0.05 \times 10^{-15}$   & \citep{meeus2012,fedele2013} \\
CI                & 1-0            & $6.60\pm 2.0 \times 10^{-19}$    & \citep{Kama2016b} \\
CI                & 2-1            & $\leq 3.58 \times 10^{-18}$      & \citep{Kama2016b} \\
CII               & 158$\mu$m      & $1.35\pm 0.15 \times 10^{-16}$   & \citep{meeus2012,fedele2013} \\
HD                & 112$\mu$m      & $\leq2.7 \times 10^{-16}$        & \citep{Kama2016b} \\
HD                & 56$\mu$m       & $\leq1.6 \times 10^{-17}$        & \citep{fedele2013} \\
C$_2$H            & 29             & $\leq4.8\times 10^{-21}$         & \citep{Kama2016b} \\
C$_2$H            & 30             & $\leq4.8\times 10^{-21}$         & \citep{Kama2016b} \\
C$_2$H            & 31             & $\leq4.8\times 10^{-21}$         & \citep{Kama2016b} \\
C$_2$H            & 32             & $\leq4.8\times 10^{-21}$         & \citep{Kama2016b} \\
C$_2$H            & 33             & $\leq4.8\times 10^{-21}$         & \citep{Kama2016b} \\
C$_2$H            & 34             & $\leq4.8\times 10^{-21}$         & \citep{Kama2016b} \\
C$_2$H            & 35             & $\leq4.8\times 10^{-21}$         & \citep{Kama2016b} \\
C$_2$H            & 36             & $\leq4.8\times 10^{-21}$         & \citep{Kama2016b} \\
C$_2$H            & 37             & $\leq4.8\times 10^{-21}$         & \citep{Kama2016b} \\
C$_2$H            & 38             & $\leq4.8\times 10^{-21}$         & \citep{Kama2016b} \\
C$_2$H            & 39             & $\leq4.8\times 10^{-21}$         & \citep{Kama2016b} \\
HCO$^+$           & 1-0            & $\leq7.0\times 10^{-20}$         & \citep{Kama2016b} \\
HCO$^+$           & 4-3            & $6.81\pm 1.0 \times 10^{-20}$    & \citep{Kama2016b} \\
H$_2$O            & 63.32          & $\leq2.59\times 10^{-17}$        & \citep{meeus2012} \\
H$_2$O            & 71.946         & $\leq5.09\times 10^{-17}$        & \citep{meeus2012} \\
H$_2$O            & 78.74          & $\leq5.70\times 10^{-17}$        & \citep{meeus2012} \\
H$_2$O            & 179.52         & $3.05\pm 4.8\times 10^{-17}$     & \citep{sturm_2010}\\
H$_2$O            & 180.42         & $\leq1.71\times 10^{-17}$        & \citep{meeus2012} \\
H$_2$O            & 90.00          & $1.160\pm 0.161 \times 10^{-16}$ & \citep{sturm_2010}\\
H$_2$O            & 557 GHz        & $1.41\pm 0.0259 \times 10^{-18}$ & \citep{Du_2017} \\
H$_2$O            & 1113 GHz       & $5.09\pm 0.0844 \times 10^{-18}$ & \citep{Du_2017} \\
H$_2$O            & 1153 GHz       & $\leq1.98\times 10^{-17}$        & \citep{Du_2017} \\
SO                & 7$_7$-6$_6$          & $1.25 \times 10^{-21}$     & \citep{booth_2022} \\
SO                & 7$_8$-6$_7$          & $1.45 \times 10^{-21}$     & \citep{booth_2022} \\
C$^{18}$O         & 3-2                       & 9.03$\times 10^{-20}$         & This work \\
SO                & 2$_1$-1$_0$               & $\leq 7.25\times 10^{-21}$    & This work \\
SO$_2$            & 4$_{3,1}$-3$_{2,2}$       & $\leq 5.32\times 10^{-21}$    & This work \\
C$^{36}$S         & 7-6                       & $\leq 5.10\times 10^{-21}$    & This work \\
HCS$^+$           & 8-7                       & $\leq 8.08\times 10^{-21}$    & This work \\
CS                & 7-6                       & 1.167$\times 10^{-20}$        & This work \\
H$_2$CS           & 10$_{0,10}$-9$_{0,9}$     & $\leq 1.72\times 10^{-20}$    & This work \\
$^{34}$SO         & 2$_3$-2$_1$               & $\leq 1.84\times 10^{-20}$    & This work \\
SO                & 8$_8$-7$_7$               & $\leq 1.72\times 10^{-20}$    & This work \\
\toprule
\end{tabular}}
\end{table*}

\twocolumn

\clearpage

\end{document}